\documentclass[useAMS,usenatbib]{mn2e}

\RequirePackage{xspace}
\RequirePackage{amsmath}
\RequirePackage{amsfonts}
\RequirePackage{amssymb}

\def\ifundefined#1{\expandafter\ifx\csname#1\endcsname\relax}

\def\la{\mathrel{\hbox{\rlap{\hbox{\lower4pt\hbox{$\sim$}}}\hbox{$<$}}}}
\def\ga{\mathrel{\hbox{\rlap{\hbox{\lower4pt\hbox{$\sim$}}}\hbox{$>$}}}}

\newcommand{\be}{\begin{equation}}
\newcommand{\ee}{\end{equation}}

\newcommand{\bea}{\begin{eqnarray}}
\newcommand{\eea}{\end{eqnarray}}

\ifundefined{ensuremath}\def\ensuremath#1{\relax\ifmmode{#1}}
\else${#1}$\fi\else\relax\fi
\ifundefined{nuc}\def\nuc#1#2{\relax\ifmmode{}^{#1}{\protect\text{#2}}
\else${}^{#1}$#2\fi}\else\relax\fi

\usepackage{graphicx}
\usepackage[T1]{fontenc} 
\usepackage{aecompl}
\usepackage{aas_macros}

\graphicspath{{./figs/}}
\def\altaffilmark#1{$^{#1}$}
\def\altaffiltext#1#2{$^{#1}$#2}
\newcounter{aaffilcoun}

\newcounter{affilcoun}

\usepackage{xargs}                      
\newcommand{\RNum}[1]{\uppercase\expandafter{\romannumeral #1\relax}}
\newcommand\ion[2]{#1$\;${\small\rmfamily\RNum#2}\relax}\title[Late-time spectra of SN~2011fe]{Optical and ultraviolet spectroscopic analysis of SN~2011fe at late times}

\author[Friesen et al.]{Brian~Friesen\altaffilmark{1,2},
        E.~Baron\altaffilmark{1,3,2},
        Jerod~T.~Parrent\altaffilmark{4},
        R.~C.~Thomas\altaffilmark{2},
        David~Branch\altaffilmark{1},
\newauthor
        Peter~Nugent\altaffilmark{2},
        Peter~H.~Hauschildt\altaffilmark{3},
        Ryan~J.~Foley\altaffilmark{5,6},
        Darryl~E.~Wright\altaffilmark{7},
\newauthor
        Yen-Chen Pan\altaffilmark{5},
        Alexei~V.~Filippenko\altaffilmark{8},
        Kelsey~I.~Clubb\altaffilmark{8},
        Jeffrey~M.~Silverman\altaffilmark{9,16},
\newauthor
        Keiichi~Maeda\altaffilmark{10,11},
        Isaac~Shivvers\altaffilmark{8},
        Patrick~L.~Kelly\altaffilmark{8},
        Daniel~P.~Cohen\altaffilmark{8},
\newauthor
        Armin~Rest\altaffilmark{12,13},
        Daniel~Kasen\altaffilmark{8,14,15}\\
        \altaffiltext{1}{Homer L. Dodge Department of Physics \& Astronomy, 440 W. Brooks St., Rm 100, Norman, OK 73019, USA}\\
        \altaffiltext{2}{Computational Research Division, Lawrence Berkeley National Laboratory, 1 Cyclotron Road MS 50B-4206, Berkeley, CA 94720, USA}\\
        \altaffiltext{3}{Hamburger Sternwarte, Gojenbergsweg 112, 21029 Hamburg, Germany}\\
        \altaffiltext{4}{Harvard-Smithsonian Center for Astrophysics, 60 Garden Street, Cambridge, MA 02138, USA}\\
        \altaffiltext{5}{Astronomy Department, University of Illinois at Urbana-Champaign, 1002 West Green Street, Urbana, IL 61801, USA}\\
        \altaffiltext{6}{Department of Physics, University of Illinois Urbana-Champaign, 1110 West Green Street, Urbana, IL 61801, USA}\\
        \altaffiltext{7}{Astrophysics Research Centre, School of Mathematics and Physics, Queen's University Belfast, Belfast BT7 1NN, UK}\\
        \altaffiltext{8}{Department of Astronomy, University of
          California, 501 Campbell Hall \#3411, Berkeley, CA 94720-3411, USA}\\
                \altaffiltext{9}{Department of Astronomy, University of Texas at Austin, Austin, TX 78712, USA}\\
        \altaffiltext{10}{Department of Astronomy, Kyoto University, Kitashirakawa-Oiwake-cho, Sakyo-ku, Kyoto 606-8502, Japan}\\
        \altaffiltext{11}{Kavli Institute for the Physics and Mathematics of the Universe, University of Tokyo, 5-1-5 Kashiwanoha, Kashiwa, Chiba 277-8583, Japan}\\
        \altaffiltext{12}{Department of Physics, Harvard University, 17 Oxford Street, Cambridge, MA 02138, USA}\\
        \altaffiltext{13}{Space Telescope Science Institute, 3700 San Martin Drive, Baltimore, MD 21218, USA}\\
        \altaffiltext{14}{Department of Physics, University of California, 366 LeConte Hall MC 7300, Berkeley, CA 94720-7300 USA}\\
        \altaffiltext{15}{Nuclear Science Division, Lawrence Berkeley National Laboratory, 1 Cyclotron Road, Berkeley, CA 94720, USA}\\
        \altaffiltext{16}{NSF Astronomy and Astrophysics Postdoctoral Fellow}\\}

\begin{document}

\date{Accepted xxx Received xx; in original form xxx}

\pagerange{\pageref{firstpage}--\pageref{lastpage}} \pubyear{2016}

\maketitle

\label{firstpage}

\begin{abstract}
  We present optical spectra of the nearby Type~Ia supernova SN~2011fe
  at 100, 205, 311, 349, and 578 days post-maximum light, as well as
  an ultraviolet spectrum obtained with \textit{Hubble Space
    Telescope} at 360 days post-maximum light.  We compare these
  observations with synthetic spectra produced with the radiative
  transfer code \texttt{PHOENIX}.  The day +100 spectrum can be well
  fit with models which neglect collisional and radiative data for
  forbidden lines.  Curiously, including this data and recomputing the
  fit yields a quite similar spectrum, but with different combinations
  of lines forming some of the stronger features.  At day +205 and
  later epochs, forbidden lines dominate much of the optical spectrum
  formation; however, our results indicate that recombination, not
  collisional excitation, is the most influential physical process
  driving spectrum formation at these late times.  Consequently, our
  synthetic optical and UV spectra at all epochs presented here are
  formed almost exclusively through recombination-driven fluorescence.
  Furthermore, our models suggest that the ultraviolet spectrum even
  as late as day +360 is optically thick and consists of permitted
  lines from several iron-peak species.  These results indicate that
  the transition to the ``nebular'' phase in Type~Ia supernovae is
  complex and highly wavelength-dependent.
\end{abstract}

\section{Introduction}
\label{sec:intro}

The Type~Ia supernova (SN~Ia) SN~2011fe was discovered on 2011 August 24, just 11~hr after explosion \citep{nugent11}.
It is among the nearest ($\sim 6.9$~Mpc) and youngest ($\sim 11$~hr) SNe~Ia ever discovered.
Extensive spectroscopic and photometric studies of SN~2011fe indicate that it is ``normal'' in nearly every sense: in luminosity, spectral and color evolution, abundance patterns, etc. \citep{parrent12,richmond12,roepke12,vinko12,munari13,pereira13}.
Its unremarkable nature coupled with the wealth of observations made over its lifetime render it an ideal laboratory for understanding the physical processes which govern the evolution of normal SNe~Ia.
Indeed, these data have allowed observers to place numerous and unprecedented constraints on the progenitor system of a particular SN~Ia \citep[e.g.,][]{li11,nugent11,bloom12,chomiuk12,horesh12,margutti12}.

Equally as information-rich as observations taken at early times are those taken much later, when the supernova's photosphere has receded and spectrum formation occurs deep in the SN core.
For example, \citet{shappee13} used late-time spectra to further constrain the progenitor system of SN~2011fe, namely that the amount of hydrogen stripped from the putative companion must be $< 0.001~M_\odot$.
\citet{mcclelland13} found that the luminosity from SN~2011fe in the 3.6~$\mu$m channel of \textit{Spitzer}/IRAC fades almost twice as quickly as in the 4.5~$\mu$m channel, which they argue is a consequence of recombination from doubly ionized to singly ionized iron peak elements.
In addition, \citet{kerzendorf14} used photometric observations near 930~d post-maximum light to construct a late-time quasi-bolometric light curve, and showed that the luminosity continues to trace the radioactive decay rate of $^{56}$Co quite closely, suggesting that positrons are fully trapped in the ejecta, disfavoring a radially combed or absent magnetic field in this SN.
\citet{graham15} presented an optical spectrum at 981~d post-explosion and used constraints on both the mass of hydrogen as well as the luminosity of the putative secondary star as evidence against a single-degenerate explosion mechanism.
\citet{taubenberger15} presented an optical spectrum at 1034~d post-explosion, and speculated about the presence of [\ion{O}{1}] lines near 6300~\AA, which, if confirmed, would provide strong constraints on the mass of unburned material near the center of the white dwarf progenitor of SN~2011fe.
Non-detections of the H$\alpha$ line at both of these very late epochs also strengthened the constraints on the presence of hydrogen initially posed by \citet{shappee13}.
Finally, \citet{mazzali15} used spectrum synthesis models of SN~2011fe from 192 to 364 days post-explosion to argue for a large central mass of stable iron and a small mass of stable nickel -- about 0.23~$M_\odot$ and 0.01~$M_\odot$, respectively.

We complement these various late-time analyses with a series of radiative transfer models corresponding to a series of optical and ultraviolet (UV) spectra of SN~2011fe.

\section{Observations}
\label{sec:obs}

\begin{table}
\begin{tabular}{lll}
UT Date & Phase & Telescope \\
& (days) & $+$Instrument \\
2011 Dec 19 & $+$100 & WHT$+$ISIS \\
2012 Apr 2 & $+$205 & Lick 3-m$+$KAST \\
2012 Jul 17 & $+$311 & Lick 3-m$+$KAST \\
2012 Aug 23 & $+$349 & Lick 3-m$+$KAST \\
2013 Apr 8 & $+$578 &Lick 3-m$+$KAST 
\end{tabular}
\caption{Observing log of spectra that appear here for the first
  time. The phase is with respect to maximum light.}
\label{tab:obs}
\end{table}

\begin{figure*}
\centering
\includegraphics[scale=0.5]{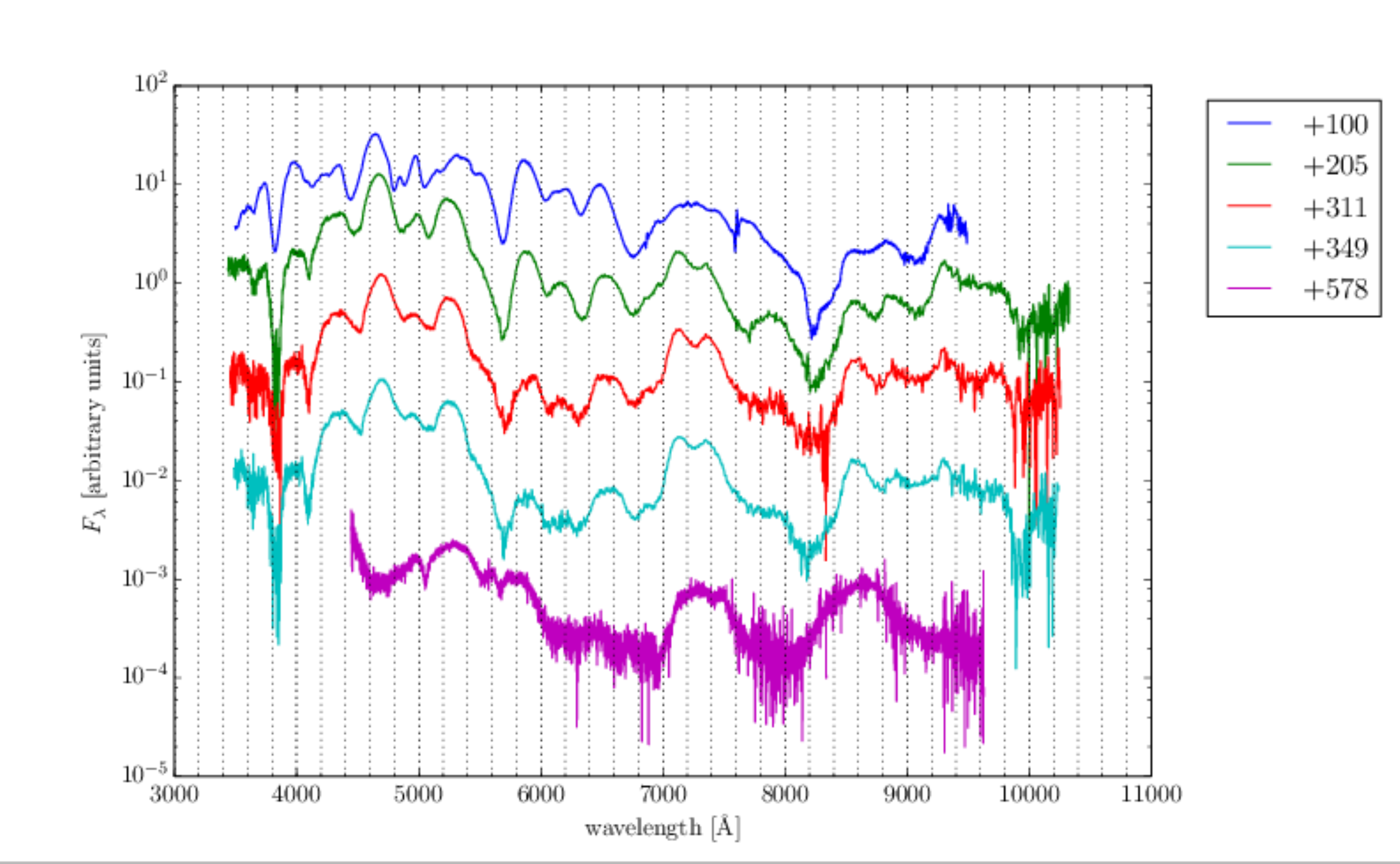}
\caption{Five optical spectra of SN~2011fe used in this work.
The fluxes are scaled arbitrarily in order to facilitate spectral feature comparisons.\label{fig:all_optical_spectra_11fe}
}
\end{figure*}

\begin{figure*}
\centering
    \includegraphics[scale=0.5]{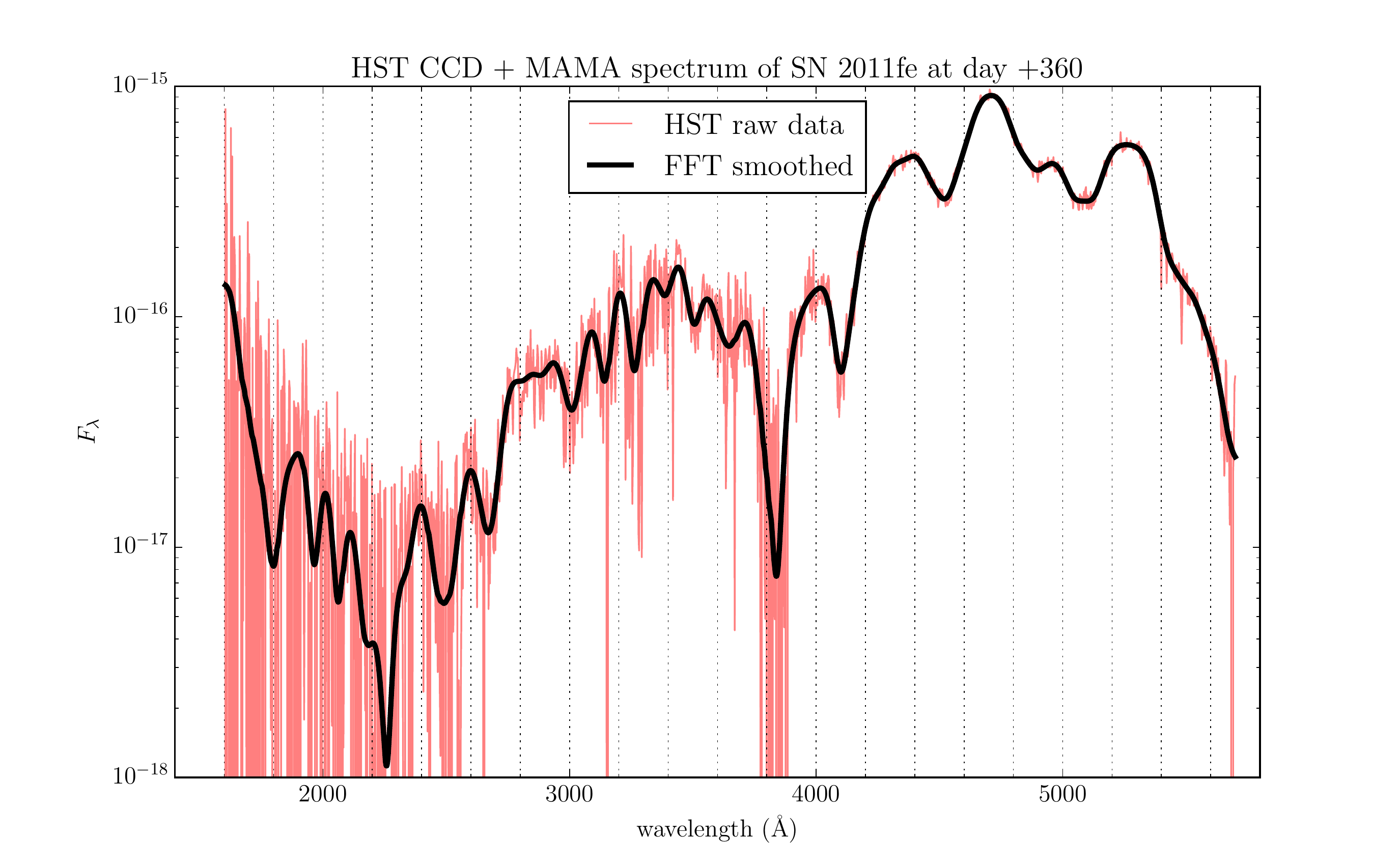}
    \caption{Smoothed spectrum using the low-pass filter technique outlined in \citet{marion09}.}
    \label{fig:sn11fe_combined_hst_spectra_ft_smoothed_no_model}
\end{figure*}

We obtained optical spectra of SN~2011fe at days +100, +205, +311,
+349, and +594 (Dec 19, 2011, Apr 2, 2012, Jul 17, 2012, Aug 23, 2012,
Mar 27, 2013); the observations are shown in
Figure~\ref{fig:all_optical_spectra_11fe} and described in
Table~\ref{tab:obs}.
The day +205 and +311 spectra were presented in \citet{mazzali15}.
We also obtained an ultraviolet spectrum with \textit{Hubble Space Telescope} at day +360 (GO 12948; PI: R.~Thomas).
This latter observation consisted of ten orbits, the first nine using the STIS/NUV-MAMA configuration, and the last with the STIS/CCD G430L and G750L configurations.
The data from one of the NUV-MAMA orbits was unrecoverable, and so the final spectrum, shown in Figure~\ref{fig:sn11fe_combined_hst_spectra_ft_smoothed_no_model}, represents co-addition of the nine remaining observations.

Also shown in Figure~\ref{fig:sn11fe_combined_hst_spectra_ft_smoothed_no_model} is a smoothed version of the \textit{HST} spectrum.
We used the algorithm presented in \citet{marion09}, which consists of applying a low-pass filter to the signal.
The motivation for this approach is the notion that the physical features in SN~Ia spectra are broad, while most noise in the spectrum is narrow.
Therefore, if one can suppress the high-``frequency'' features (the noise), what will remain will be the pure signal from the SN.
To accomplish this task, one calculates the power spectrum of the original spectrum using a Fourier transformation, suppresses the power spectrum at all high ``frequencies'' in which information is deemed to be noise, and applies an inverse Fourier transformation to recover the smoothed spectrum.
An especially useful feature of this technique is its insensitivity to spikes in \textit{HST} spectra due to cosmic rays.

\section{Radiative transfer models}

We used the \texttt{PHOENIX/1D} code \citep{hb99} with the same modifications discussed in \citet{friesen14} to capture the most important physical processes at late times in SNe~Ia. The underlying explosion model was a spherically symmetric delayed-detonation model presented in \citet{dominguez01}. For each observation presented in \S\ref{sec:obs} we calculated a corresponding synthetic spectrum, assuming a 16~day rise time for the model \citep{nugent11}.

\section{Discussion}
\label{sec:discussion}

The theory of spectrum formation at late times in normal SNe~Ia has broadly converged to a scenario in which electron configurations of atoms in the rarefied ejecta are primarily in their ground state, and are excited by collisions with free electrons to low-lying metastable levels, which in turn emit forbidden lines as they return to the ground state \citep[e.g.,][]{meyerott78,meyerott80,axelrod80,ruiz-lapuente92,kuchner94,ruiz-lapuente95,bowers97,mazzali11,mazzali12,silverman13}.
Little to no continuum emission is present in SN~Ia spectra at these epochs.
This stands in contrast to the spectrum formation mechanism at early times, near maximum light, in which the optical depth to Thomson scattering on free electrons is large, leading to the formation of a photosphere on top of which atoms undergo line scattering via strong permitted lines, giving rise to P~Cygni spectral features.
\citet{kirshner75} argued that the P~Cygni mechanism is no longer active at late times because the photosphere has disappeared (indicated by the absence of continuum) and there are no longer enough photons for these strong lines to scatter.

Curiously, good spectral fits have been obtained for relatively late SN~Ia spectra with the parameterized spectrum synthesis code \texttt{SYNOW}, which treats only line scattering by permitted lines.
Such fits require only a few ions -- \ion{Na}{1}, \ion{Ca}{2}, and
\ion{Fe}{2} -- and fit optical spectra fairly well, especially blueward of $\sim 6000$~\AA.
Examples include the normal SN~1994D at day +115 \citep{branch05}, the normal SN~2003du at day +84 \citep{branch08}, the subluminous SN~1991bg at day +91 \citep{branch08}, and the peculiar SN~2002cx at day +227 \citep{jha06}.
While the parameterized approach of \texttt{SYNOW} to solving the radiative transfer equation restricts analysis of those fits to putative line identifications and velocity measurements, they nevertheless demonstrate that either SN~Ia spectra exhibit a remarkable degeneracy with respect to forbidden and permitted line formation, or that permitted lines continue to drive emergent spectrum formation at late times, regardless of what physical mechanisms generate the underlying flux \citep{branch08,friesen12}.

These two competing analyses of late-time SN~Ia spectra agree that the majority of the spectrum is formed by Fe lines, but they predict dramatically different velocities of the line-forming regions in the ejecta.
For example, \citet{branch08} argue that Fe, Ca, and Na are located at 7000~km~s$^{-1}$ and higher in the day +84 spectrum of SN~2003du.
In contrast, \citet{bowers97} argue for velocities from 1000~--~3000~km~s$^{-1}$ in the +95 spectrum of the same object.
Identifying the correct velocity of the line-forming region has important consequences for constraining the structure of the inner regions of SN~Ia ejecta, which in turn constrain the as-yet unknown explosion mechanism.

\subsection{Day +100}
\label{subsec:11fe_p100}

\begin{figure*}
\centering
    \includegraphics[scale=0.5]{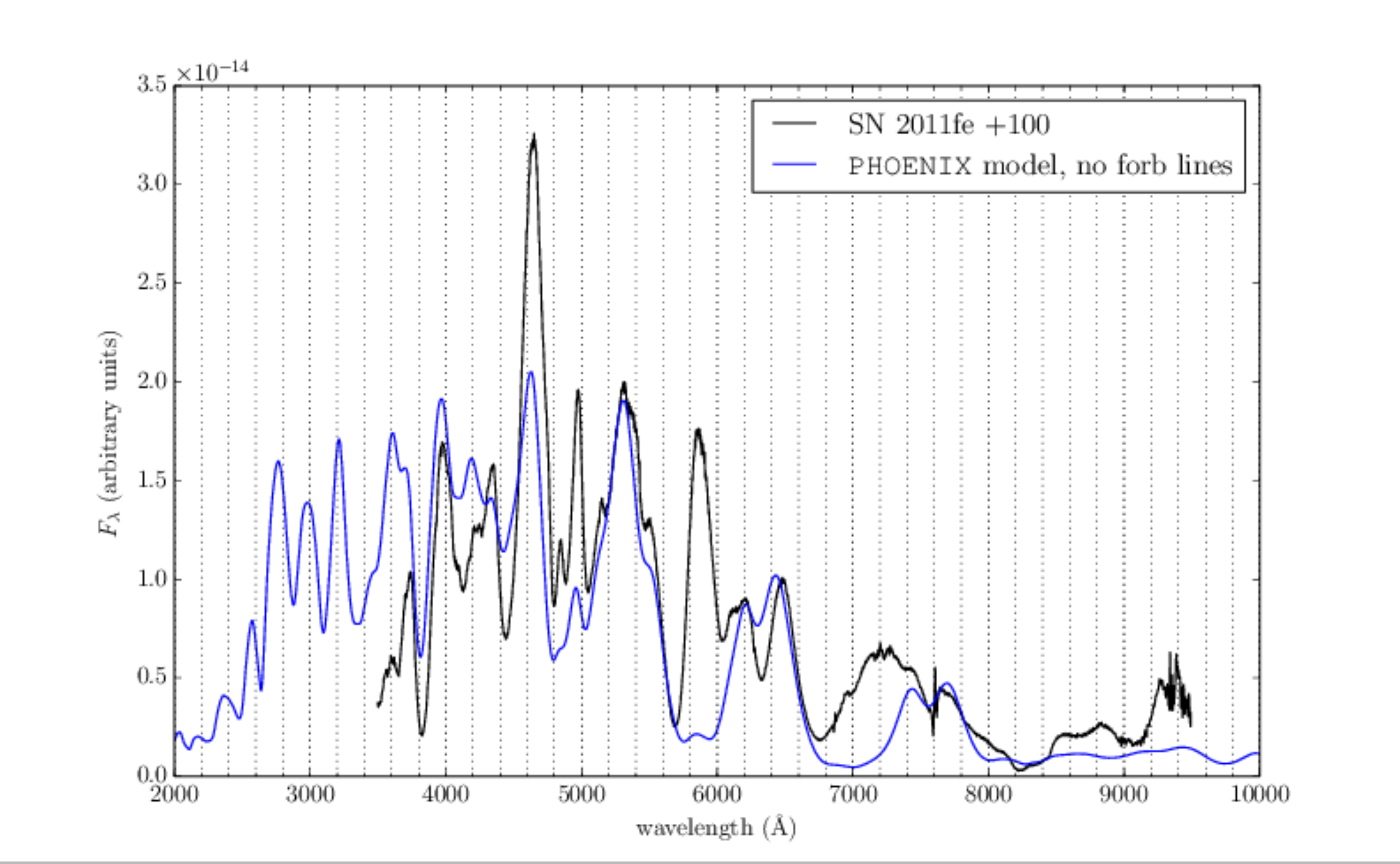}
    \caption{Synthetic spectrum of delayed-detonation model of \citet{dominguez01} at day 116 vs. SN~2011fe at day +100.
             No forbidden lines were included in this calculation.}
    \label{fig:pah_std_d116_delta_t_no_forb_lines_vs_11fe_p100}
\end{figure*}

The day +100 spectrum of SN~2011fe and the corresponding synthetic spectrum from \texttt{PHOENIX} are shown in Figure~\ref{fig:pah_std_d116_delta_t_no_forb_lines_vs_11fe_p100}.
Overall the fit is good, although a few features in our model do not match those in the observed spectrum, namely the emission feature near 5900~\AA.
In addition, the peak at 4700~\AA\ in the synthetic spectrum is too weak, the blue side of the broad emission at 7200~\AA\ is absent in the model, and the flux in the blue and near-UV is too high.
Most other features are well reproduced, both in strength and in shape.

In the context of much of the literature which concerns late-time SN~Ia spectra, the fidelity of this fit is peculiar because the calculation used the most current atomic database of \citet{kurucz95,kurucz02}, which includes no forbidden line data for any ions.
This stands in contrast to the most common interpretations of spectra of ``old'' SN~Ia, which were discussed earlier.
If the purely permitted line identifications are correct they
are difficult to reconcile with kinematic analyses such as that of \citet{maeda10}, which assume that the emission peaks correspond to forbidden lines forming within a few 100~km~s$^{-1}$ of $v \simeq 0$~km~s$^{-1}$.
Rather, the strong emission peaks at $4700$~\AA\ and $5200$~\AA, which
each have previously been identified as a mixture of [\ion{Fe}{2}] and
[\ion{Fe}{3}] lines, may instead be formed by the handful of permitted
lines of \ion{Fe}{2} whose upper levels are among the crowded
$3d^6(^5D)4p$ configuration, with energies between 5~eV and 6~eV, and
whose lower levels are, coincidentally, the handful of metastable
levels around 3~eV which are purportedly responsible for the
aforementioned forbidden emission features. However, it is important
to note that the analysis of \citet{maeda10} only requires that the
ejecta is optically thin at the rest wavelength of the line and as we
show below in the redder parts of the spectra that condition is met at
later epochs. 
In the line-scattering interpretation of spectrum formation, this
would imply that the dips just to the blue of these two strong
emission features are the corresponding absorptions, rather than
regions lacking in emission. 
These absorptions would correspond to line velocities of $\sim
6000$~km~s$^{-1}$, similar to that found in the +115 spectrum of
SN~1994D by \citep{branch05}. 

Although below we illustrate some complications with this permitted-line-only model, it is instructive first to entertain the idea that this is, in fact, representative of late-time spectrum formation physics in SNe~Ia.
Given the contrast of this result with those found elsewhere in the literature, it is important to evaluate the late-time line scattering scenario within the context of other analyses of SN~2011fe, in order to determine whether or not it is copacetic with what is already known about the spectral evolution of this object.
We consider three such pieces of analysis.
First, \citet{parrent12} traced the velocity evolution of \ion{Fe}{2} in the early spectra of SN~2011fe using the automated spectrum code \texttt{SYNAPPS} \citep{thomas11} and found that at day +15 the minimum velocity of that ion was $\sim 8000$~km~s$^{-1}$ (see their Figure~3).
Furthermore, after maximum light, the rate of change of line velocities in SN~2011fe, and in most SNe~Ia in general, slows dramatically.
Second, in the hydrodynamical model used in our calculation, Fe remains the most abundant species in the ejecta from the center of the ejecta out to $\sim 12 000$~km~s$^{-1}$ \citep[see][Figure~2]{friesen14}; our putative line velocity estimate of $\sim 6000$~km~s$^{-1}$ falls well within this range.
Finally, \citet{iwamoto99} show that the optical depth of \ion{Fe}{2} computed in local thermodynamic equilibrium (LTE) peaks between 5000~K and $10 000$~K, roughly the same temperature range as that of the ejecta in our models.
(One would be remiss to read too much into this corroboration, as the radiation field in the SN~Ia ejecta at this epoch is far from LTE.)
Although none of these results offer conclusive evidence that the strong features in the +100 spectrum of SN~2011fe are indeed P~Cygni profiles, they do show that it is quite reasonable to consider that possibility.

We caution that it is unlikely that the \emph{entire} optical spectrum consists of overlapping P~Cygni line profiles due to resonance-scattering, as is the case at very early (photospheric) epochs in SNe~Ia.
\citet{branch08} attempted to fit the day +84 spectrum of SN~2003du with the resonance-scattering code \texttt{SYNOW}, and found that P~Cygni lines fit the blue part of the spectrum (blueward of 6600~\AA) quite well, but failed quite severely redward of that.
As they discuss, the likely explanation is that resonance-scattering near this epoch is very influential at blue wavelengths, but forbidden emission is prominent in the red.
(We find a similar result in our attempts to fit the optical and UV spectra at +349 and +360, respectively, which we discuss below.)
In fact, to argue that spectrum formation consists of \emph{either} resonance-scattering by optically thick permitted lines \emph{or} emission from optically thin forbidden lines is somewhat of a false dilemma, as both scenarios assume a degree of locality in the radiative transfer which is probably unphysical.
In particular, the former assumes that the source function $S$ depends only on the local mean intensity $J$, while the latter assumes that emitting lines are well separated in wavelength such that they act independently of each other.
Each of these approximations is valid in certain regimes, i.e., resonance-scattering at photospheric epochs and forbidden emission at \emph{very} late times ($> 1$~yr) and in wavelength regions far from the forest of iron-peak lines, such as the infrared, but there exists a wide range between those extremes, in which all of these effects compete to form the emergent spectrum.

\citet{bongard08} addressed this topic in detail by calculating a grid of \texttt{PHOENIX} spectra using the hydrodynamical model W7 \citep{nomoto84} at 20 days post-explosion.
They found that even at very low velocities and high optical depth ($\tau > 3$), the ``spectrum''\footnote{The radiation flux $F_\lambda$.} at those velocities is already highly distorted from that of a blackbody, due to line and continuum interactions of the radiation field with iron-peak elements deep in the core of the SN.
The ions found at higher velocities, near the photosphere, then further distort this underlying spectrum through additional absorption, emission, and line scattering, leading to an emergent spectrum containing a complicated mixture of P~Cygni, continuum, and thermal components which are difficult to disentangle.

That our spectral model which explicitly omits forbidden line data fits the day +100 spectrum of SN~2011fe reasonably well, suggests that day +100 is simply too early for collisionally excited forbidden emission to be the primary driver of spectrum formation.
It appears that line scattering processes continue to contribute significantly, even this late in the lifetime of this SN.
In short, there are many physically-motivated reasons to suspect that permitted lines play an important role in SN~Ia spectrum formation at this epoch.

\begin{figure*}
\centering
	\includegraphics[scale=0.5]{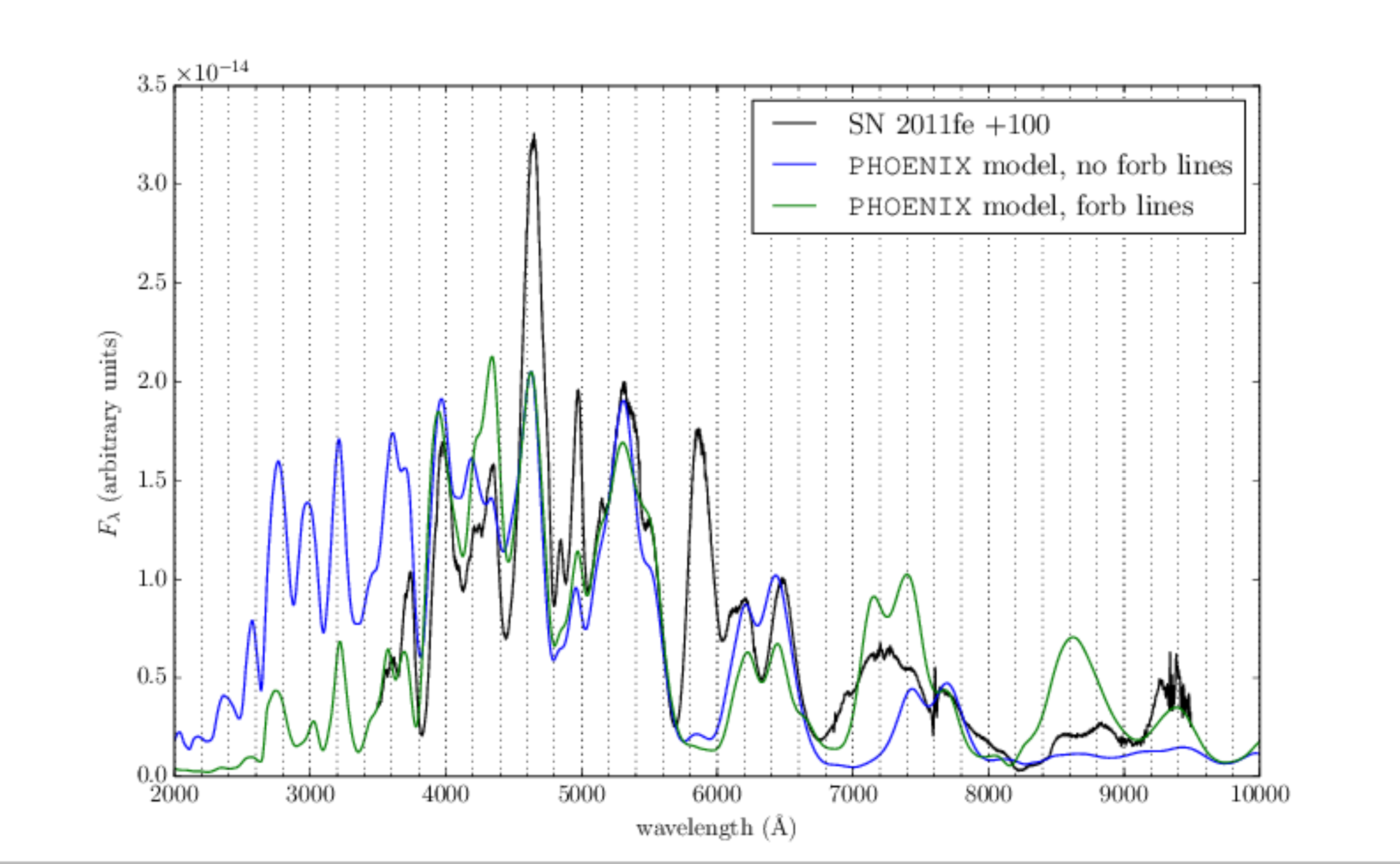}
    \caption{Comparison of \texttt{PHOENIX} spectra with and without forbidden lines at day 116.}
    \label{fig:pah_std_d116_delta_t_forb_lines_vs_no_forb_lines_vs_11fe_p100}
\end{figure*}

However, since forbidden lines are frequently identified in spectra of SNe~Ia of this age, we tested this theory by expanding our atomic database to include collisional and radiative data of forbidden lines, as described in \citet{friesen14}.
We then repeated the radiative transfer calculation with this expanded database, and compare the two results in Figure~\ref{fig:pah_std_d116_delta_t_forb_lines_vs_no_forb_lines_vs_11fe_p100}.
The results are quite similar, except that the model with forbidden lines has a lower UV flux and stronger emission at 7300~\AA\ and 8600~\AA.

The most notable shortcoming of both synthetic spectra is the lack of emission at 5900~\AA.
This feature has been identified alternatively as \ion{Na}{1}~D \citep{branch08,mazzali08} or [\ion{Co}{3}]~$\lambda 5888$~\AA\ \citep{dessart14}.
The explosion model used in these calculations contains little \ion{Na}{1}, so it is not surprising that we do not recover a strong Na~I~D emission feature there.
However, at day 116 the model contains several 0.1~$M_\odot$ of $^{56}$Co, and yet the forbidden emission at 5900~\AA\ does not appear.
This discrepancy may be related to underestimating the gas temperature in the model at this epoch (see \S\ref{subsec:11fe_p205}).

Identifying whether a feature is an ``emission'' or ``absorption'' is not a straightforward task in \texttt{PHOENIX} calculations.
This is because the algorithm calculates emissivities and opacities of NLTE species by adding up all contributions to each at each wavelength point \citep[e.g.,][]{hb14}, with no regard to the underlying atomic processes which produced them.
Such an approach captures naturally the notion that spectrum formation is inherently multi-layered in supernovae: one region deep in the ejecta may be strongly in emission at one wavelength, but a region above it may be optically thick at that wavelength, absorbing much of the underlying emission \citep{bongard08}.
The emergent spectrum is then a convolution of both processes, and such classifications as ``absorption'' or ``emission,'' while relevant to single interactions, no longer describe adequately the complete process of spectrum formation.

We are therefore relegated to using more indirect methods for isolating the sources of features in synthetic spectra.
The single-ion spectrum method \citep{bongard08} can help one identify the particular ion or ions which influence particular parts of a synthetic spectrum, but it cannot, e.g., isolate the effects of permitted lines from forbidden lines, which is desirable in this context.
We found that the only useful way to accomplish this was to remove the forbidden lines from the calculation entirely and re-compute the entire model.
This can unfortunately broaden the parameter space of the model, since forbidden lines affect the temperature structure by acting as coolants \citep{friesen14,dessart14}.
Unfortunately we are aware of no more targeted method of accomplishing this goal.

\begin{figure*}
\centering
    \includegraphics[scale=0.9]{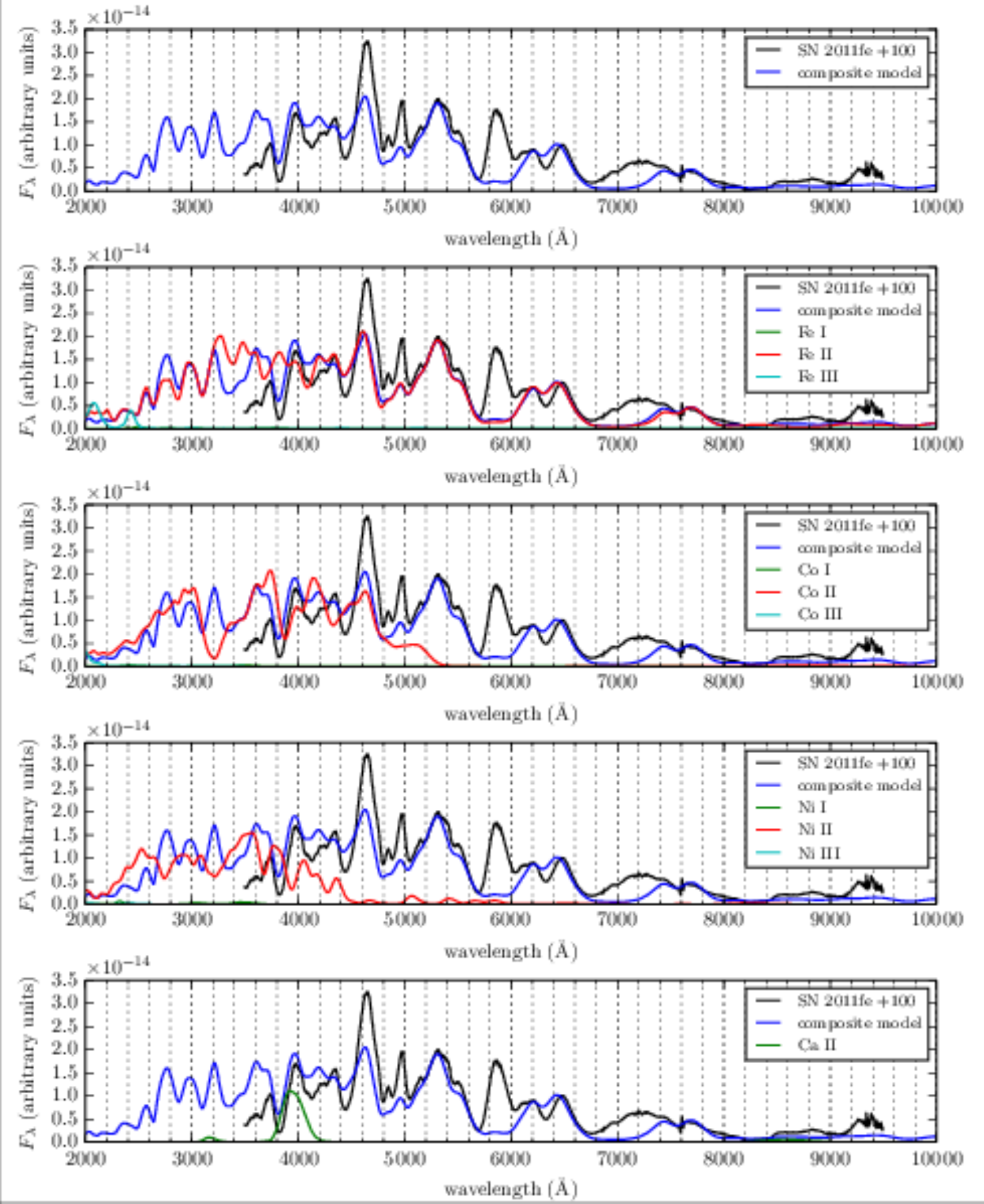}
    \caption{Single-ion spectra corresponding to the composite spectrum of the delayed-detonation model of \citet{dominguez01} at day 116, compared to SN~2011fe at day +100.
             No forbidden lines were included in this calculation.}
    \label{fig:pah_std_d116_delta_t_no_forb_lines_vs_11fe_p100_single_ion_spectra}
\end{figure*}

\begin{figure*}
\centering
    \includegraphics[scale=0.9]{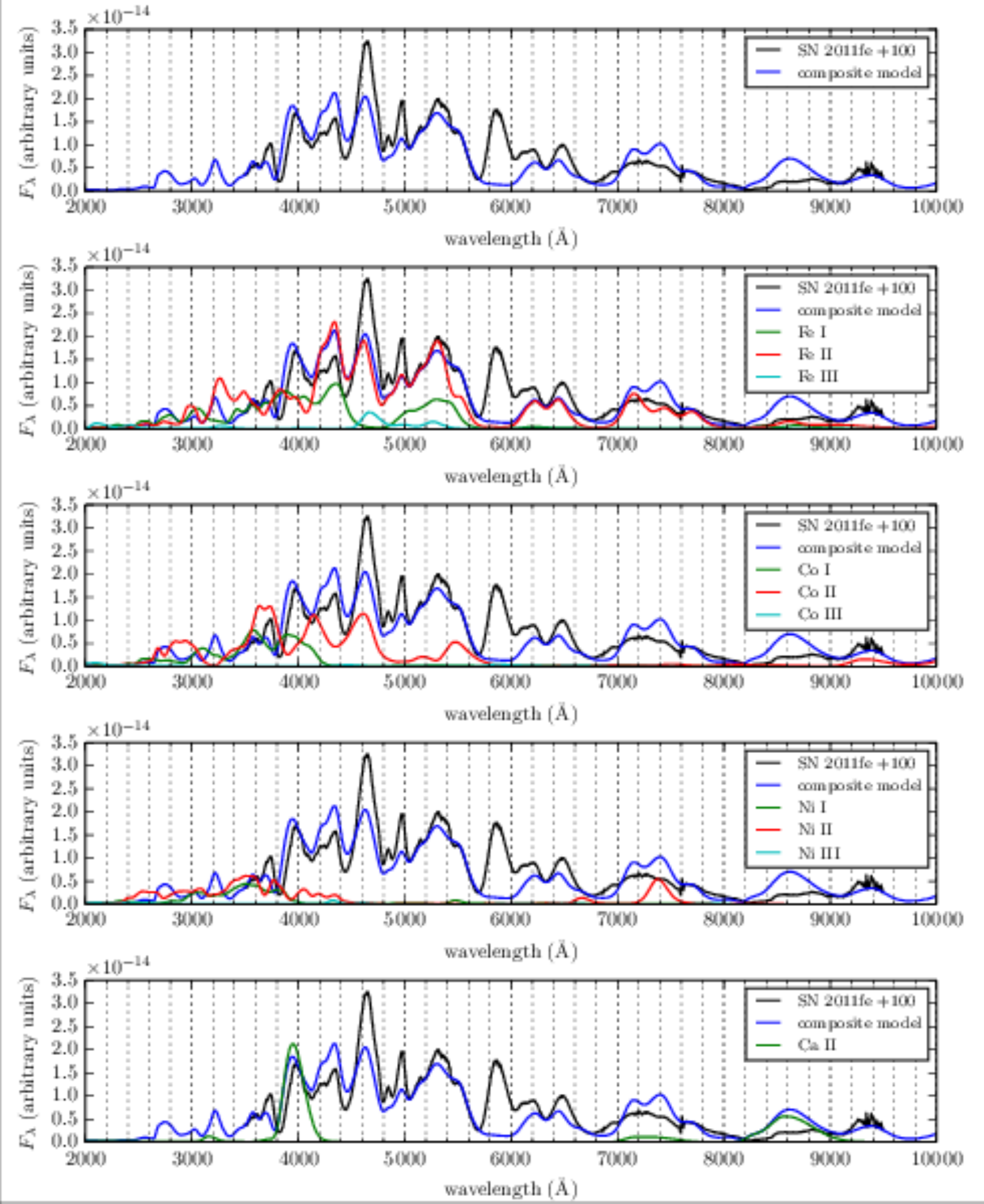}
    \caption{Single-ion spectra corresponding to the composite spectrum of the delayed-detonation model of \citet{dominguez01} at day 116, compared to SN~2011fe at day +100.}
    \label{fig:pah_std_d116_delta_t_forb_coll_vs_11fe_p100_single_ion_spectra}
\end{figure*}

We computed single-ion spectra for both synthetic spectra (with and without forbidden lines), for the most influential ions.
For the spectrum without forbidden lines, these are shown in Figure~\ref{fig:pah_std_d116_delta_t_no_forb_lines_vs_11fe_p100_single_ion_spectra}.
For the spectrum with forbidden lines, these are shown in Figure~\ref{fig:pah_std_d116_delta_t_forb_coll_vs_11fe_p100_single_ion_spectra}.

Both with and without forbidden lines, the synthetic spectra indicate that most of the optical spectrum at day +100 is formed by \ion{Fe}{2}.
In addition, the \ion{Ca}{2} H \& K doublet at $\lambda \lambda 3934, 3968$~\AA, a pair of strong resonance lines, contributes significantly to the emission at 4000~\AA.
However, comparison between the two sets of single-ion spectra indicate a fascinating result which \citet{branch08} found highly improbable: it appears that entirely different combinations of atomic lines can conspire to produce similar optical spectra.
Furthermore, the synthetic spectra in Figure~\ref{fig:pah_std_d116_delta_t_no_forb_lines_vs_11fe_p100} and Figure~\ref{fig:pah_std_d116_delta_t_forb_lines_vs_no_forb_lines_vs_11fe_p100} are the \emph{natural endpoints} of calculations subject to otherwise identical parameters.
For example, when forbidden lines are included, the emission at 4000~\AA\ is due entirely to the \ion{Ca}{2} H \& K doublet; when they are absent, it is a combination of that same doublet with contributions also from \ion{Fe}{2} and \ion{Co}{2}.
The double-horned emissions at 7250~\AA\ and 7500~\AA\ in the synthetic spectra lacking forbidden lines are due to emission from \ion{Fe}{2}; but the double-horned features at 7150~\AA\ and 7400~\AA\ in the spectra containing forbidden lines are due to \ion{Fe}{2} (possibly [\ion{Fe}{2}]~$\lambda \lambda 7155, 7171$~\AA)and [\ion{Ni}{2}]~$\lambda 7374, 7412$~\AA.
This is likely stable $^{58}$Ni, since the radioactive $^{56}$Ni has mostly decayed by this point.
The degeneracy among these various features is the likely explanation for the conflicting results of, e.g., \citet{bowers97} and \citet{branch08}.

One would expect that adding forbidden lines is always favorable over neglecting them: if a calculation captures all relevant atomic processes and if forbidden lines are truly unimportant in some scenario, they will naturally ``deactivate''.
And in fact, adding in the forbidden atomic data did address some problems in the synthetic spectra which lacked them.
The lower UV flux and the 8600~\AA\ emission in the synthetic spectrum forbidden lines can be explained by their cooling effects: lower temperatures generally lead to lower opacities in the UV, and a lower temperature allows more \ion{Ca}{3} to recombine to \ion{Ca}{2}.
However, the cooling effects also introduced a new problem: the infrared triplet (IR3) of \ion{Ca}{2} $\lambda \lambda 8498, 8542, 8662$~\AA\ (a trio of strong permitted lines) is responsible for the emission at 8600~\AA, but the model overestimates the strength of the emission at this wavelength.
In the observation there is a pair of weaker emission features at the same location, and it is possible that at least one of these two emissions is due to the \ion{Ca}{2} IR3, although probably not both, since they are spread too far apart in wavelength.
It is therefore not entirely clear which of the two synthetic spectra are ``better,'' and it is possible therefore that both permitted \emph{and} forbidden lines affect the optical spectra of SNe~Ia at this epoch.

\subsection{Day +205}
\label{subsec:11fe_p205}

\begin{figure*}
\centering
\includegraphics[scale=0.5]{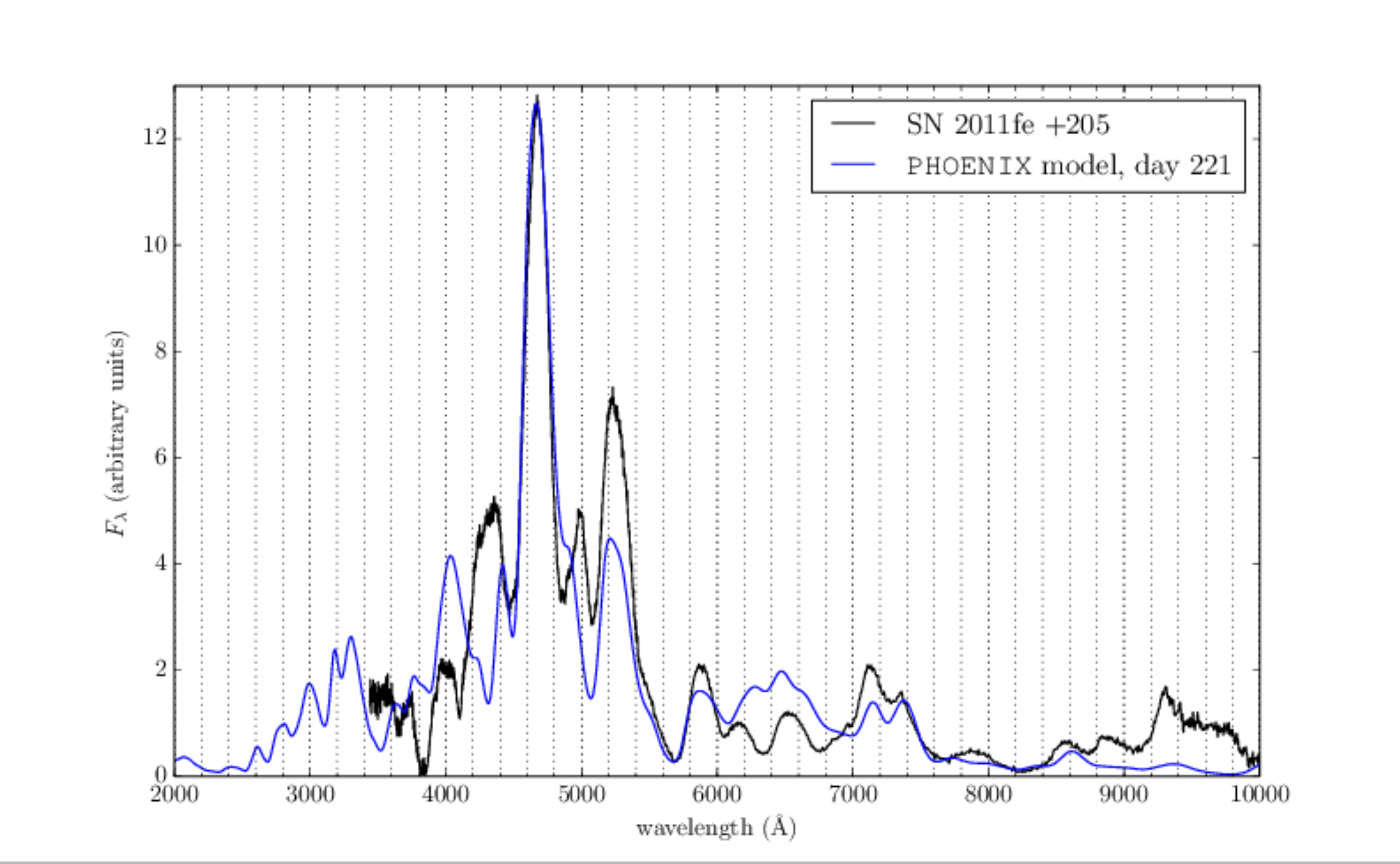}
    \caption{Synthetic spectrum of delayed-detonation model of \citet{dominguez01} at day 221 vs. SN~2011fe at day +205.}
    \label{fig:pah_std_d221_delta_t_forb_coll_vs_11fe_p205}
\end{figure*}

\begin{figure*}
  \centering
    \includegraphics[scale=0.9]{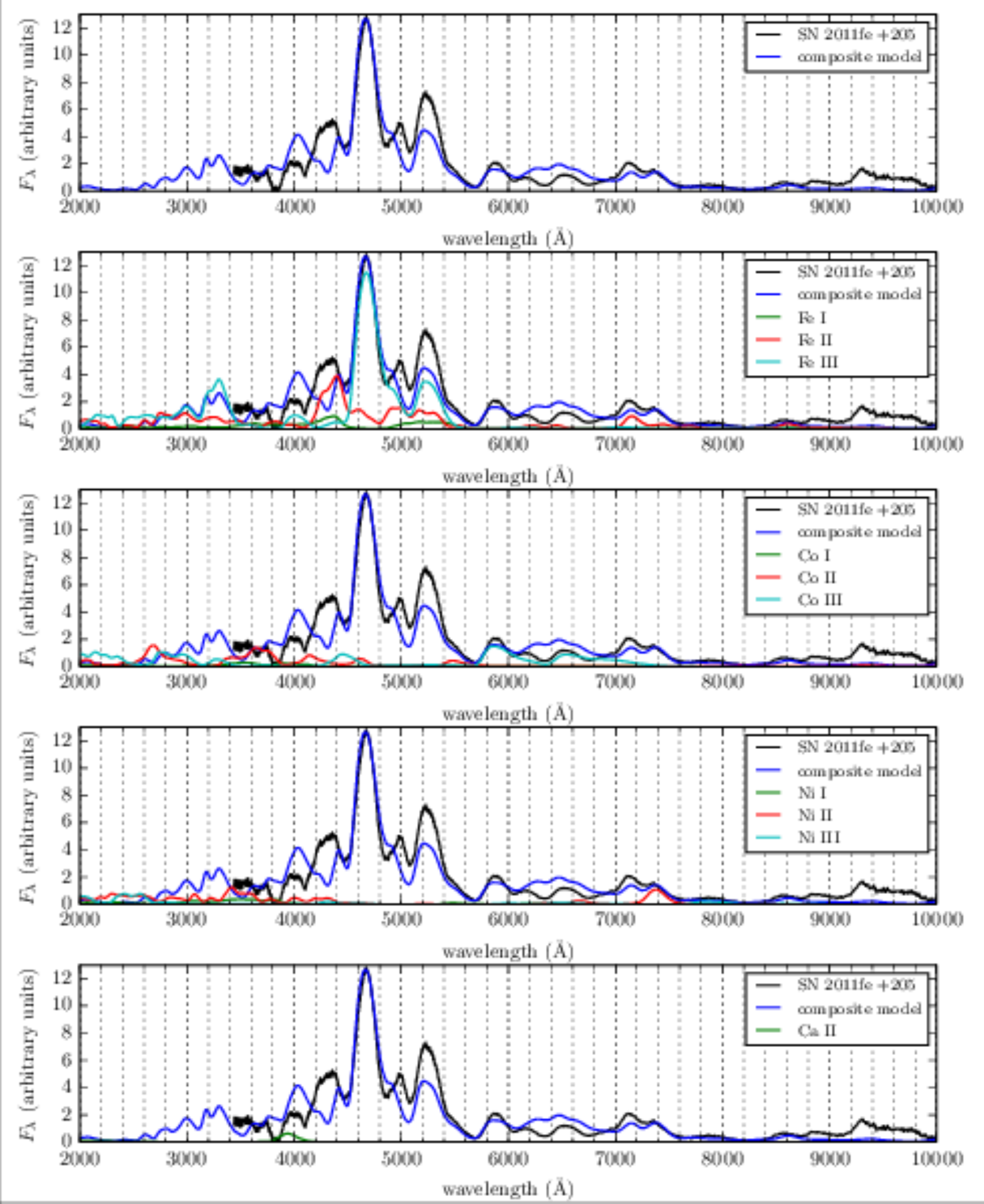}
    \caption{Single-ion spectra corresponding to the composite spectrum of the delayed-detonation model of \citet{dominguez01} at day 221, compared to SN~2011fe at day +205.}
    \label{fig:pah_std_d221_delta_t_forb_coll_vs_11fe_p205_single_ion_spectra}
\end{figure*}

The day +205 spectrum of SN~2011fe and the corresponding \texttt{PHOENIX} spectrum are shown in Figure~\ref{fig:pah_std_d221_delta_t_forb_coll_vs_11fe_p205}.
Attempts to calculate a spectrum at this epoch without forbidden lines, as was done in \S\ref{subsec:11fe_p100}, led to unrecoverable numerical instabilities in the code.
It seems, then, that by this age forbidden lines play an important role.
The single-ion spectra are shown in Figure~\ref{fig:pah_std_d221_delta_t_forb_coll_vs_11fe_p205_single_ion_spectra}.
The emission feature at 4700~\AA\ is primarily [\ion{Fe}{3}]~$\lambda \lambda 4607, 4658$~\AA\ and [\ion{Fe}{2}]~$\lambda \lambda 4640, 4664$~\AA.
The weak but clearly separate features around 4300~\AA\ are [\ion{Fe}{2}]~$\lambda \lambda 4287, 4359$~\AA, and the emission at 5300~\AA\ is [\ion{Fe}{2}]~$\lambda 5300$~\AA.
The double-horned feature in the synthetic spectrum centered around 7300~\AA\ consists of [\ion{Fe}{2}]~$\lambda \lambda 7155, 7172$~\AA\ on the left, and [\ion{Ni}{2}]~$\lambda 7412$~\AA\ on the right; the shape of this pair of features is too exaggerated in the synthetic spectrum compared to the day +100 spectrum of SN~2011fe, but at later epochs the shape is a good match to the observations.

At this epoch the ratio of \ion{Fe}{2} to \ion{Fe}{3} is well reproduced, with the strength of the 4700~\AA\ emission from \ion{Fe}{3} improved over that from the day +100 spectrum.
However, the 5200~\AA\ emission, also from \ion{Fe}{3}, is somewhat weak.
The strong \ion{Ca}{2} IR3 emission which was overestimated in strength in they day +100 synthetic spectrum is now absent entirely.
Coincidentally, the forbidden line [\ion{Fe}{2}]~$\lambda 8617$~\AA, at nearly the same wavelength as IR3, has grown in strength at day +205, and fits quite well to the observation.
The \ion{Ca}{2} H \& K doublet, which was quite strong at day +100, has diminished in strength and is replaced mostly by [\ion{Fe}{3}]~$\lambda 4008$~\AA.
It seems, then, that the serendipitous degeneracy among permitted and forbidden lines which \citet{branch05} found unlikely, may actually be realized, at least for some features in the optical spectra of SN~2011fe.

The double-horned pair of emissions centered at 7200~\AA\ is better reproduced at this epoch as well.
Curiously, the emission at 5900~\AA\ is now quite well fit with [\ion{Co}{3}]~$\lambda 5888$~\AA, while at day +100, when most of the $^{56}$Ni had decayed to $^{56}$Co, the feature was absent entirely.
It is possible that the temperature in the day +100 model was too low, which would explain the underabundance of \ion{Fe}{3} emitting at 4700~\AA\ and \ion{Co}{3} emitting at 5900~\AA.
A higher temperature would also reduce the abundance of \ion{Ca}{2} in favor of \ion{Ca}{3}, explaining the reduced strength of both the H \& K doublet and the IR3.

\subsection{Day +311}

\begin{figure*}
\centering
    \includegraphics[scale=0.5]{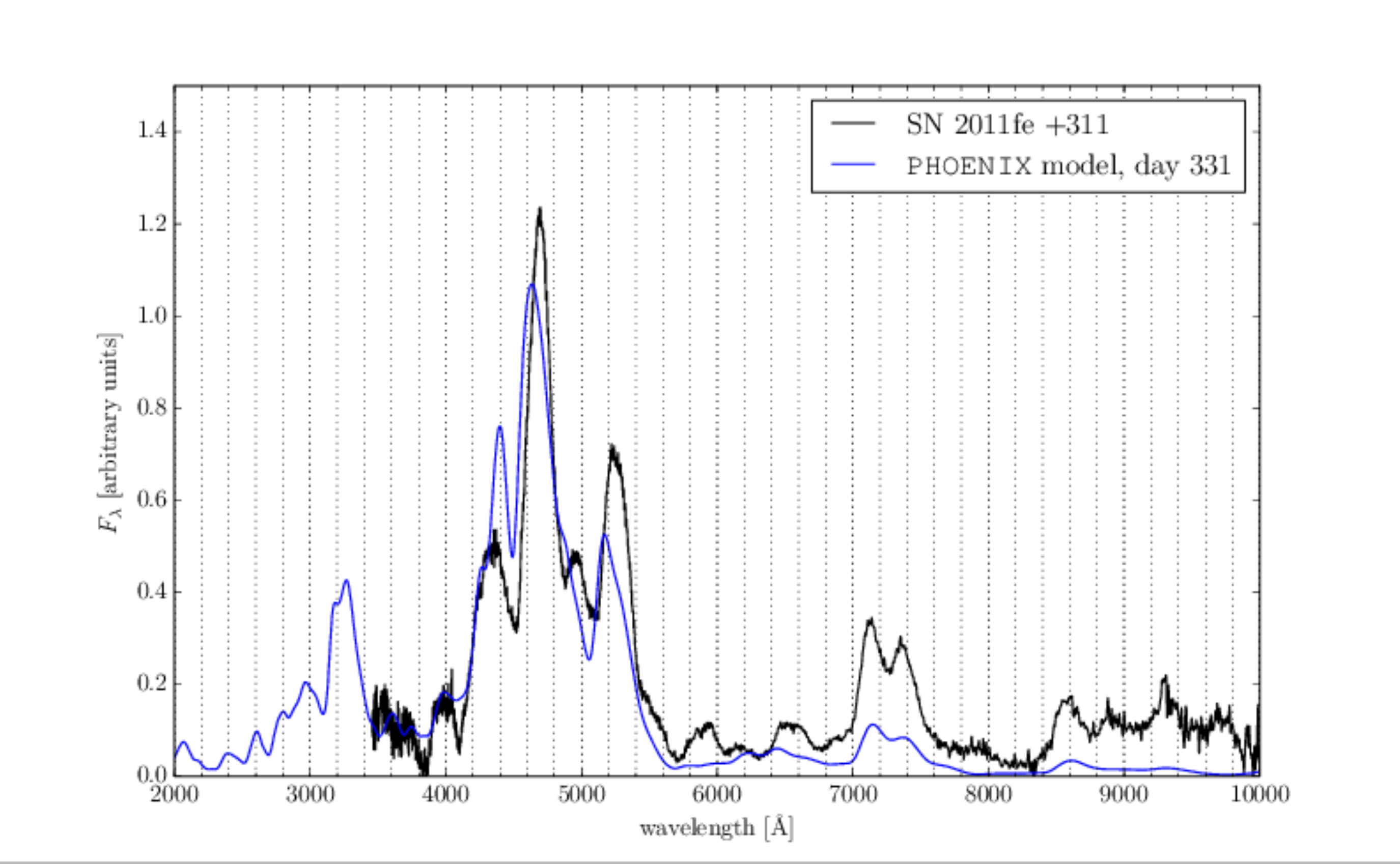}
    \caption{Synthetic spectrum of delayed-detonation model of \citet{dominguez01} at day 331 vs. SN~2011fe at day +311.}
    \label{fig:pah_std_d331_delta_t_forb_coll_vs_11fe_p311}
\end{figure*}

\begin{figure*}
\centering
\includegraphics[scale=0.7]{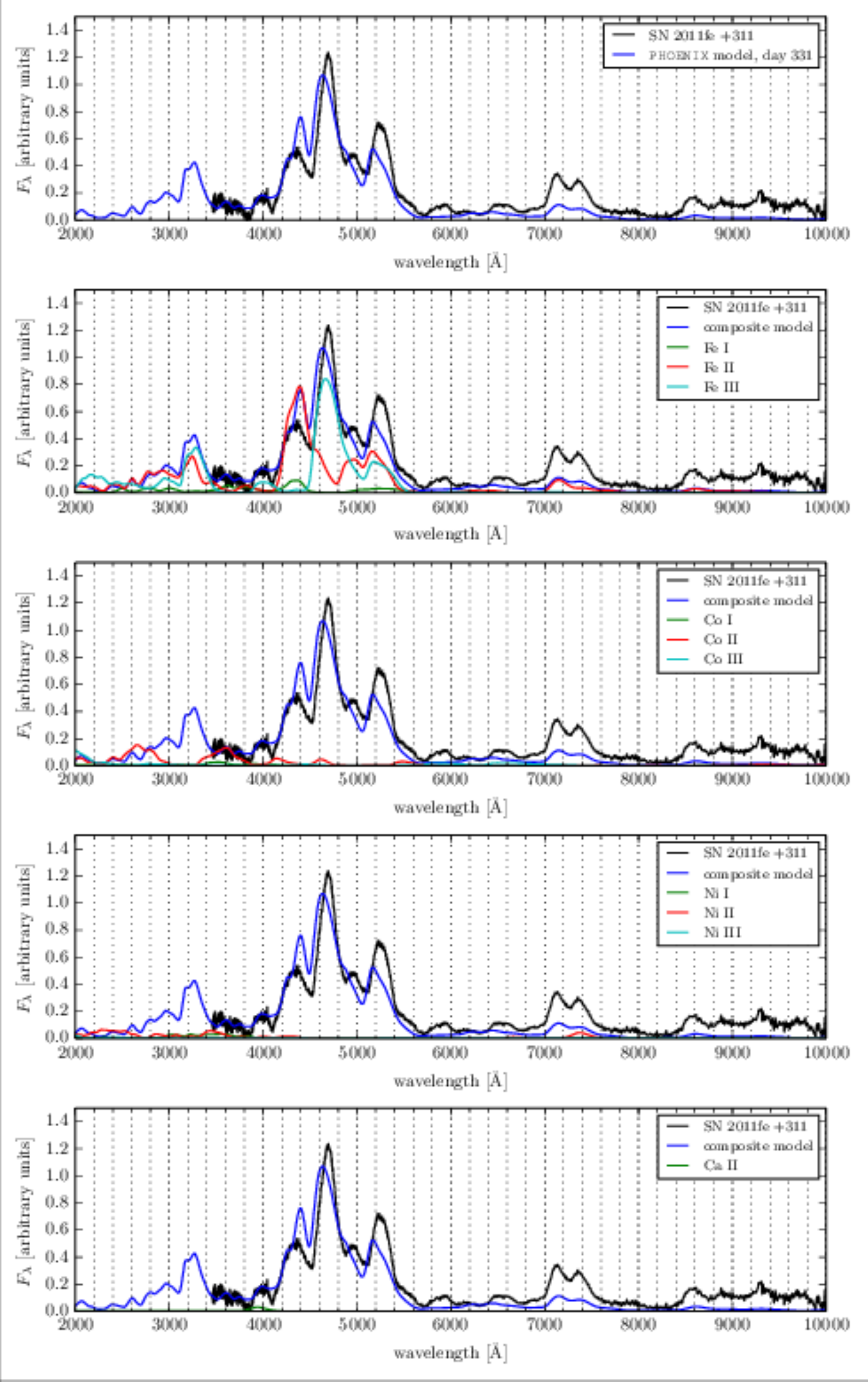}
    \caption{Single-ion spectra corresponding to the composite spectrum of the delayed-detonation model of \citet{dominguez01} at day 331, compared to SN~2011fe at day +311.}
    \label{fig:pah_std_d331_delta_t_forb_coll_vs_11fe_p311_single_ion_spectra}
\end{figure*}

The observed and synthetic spectra at day +311 are shown in Figure~\ref{fig:pah_std_d331_delta_t_forb_coll_vs_11fe_p311}, and the single-ion spectra are shown in Figure~\ref{fig:pah_std_d331_delta_t_forb_coll_vs_11fe_p311_single_ion_spectra}.
At this epoch the spectra look similar to those at day +205.
The [\ion{Fe}{3}] emission at 4700~\AA\ is still strong, although at day +311 the [\ion{Fe}{2}] emission at 4400~\AA\ has grown in strength, and continues to do so at later epochs.
This is likely a reflection of some (but not much) recombination from \ion{Fe}{3} to \ion{Fe}{2} at this age.
The \ion{Ca}{2} H \& K emission is still present at 4000~\AA, but somewhat weak, just as at day +205.

\subsection{Day +349}
\label{subsec:11fe_p349}

\begin{figure*}
\centering
    \includegraphics[scale=0.5]{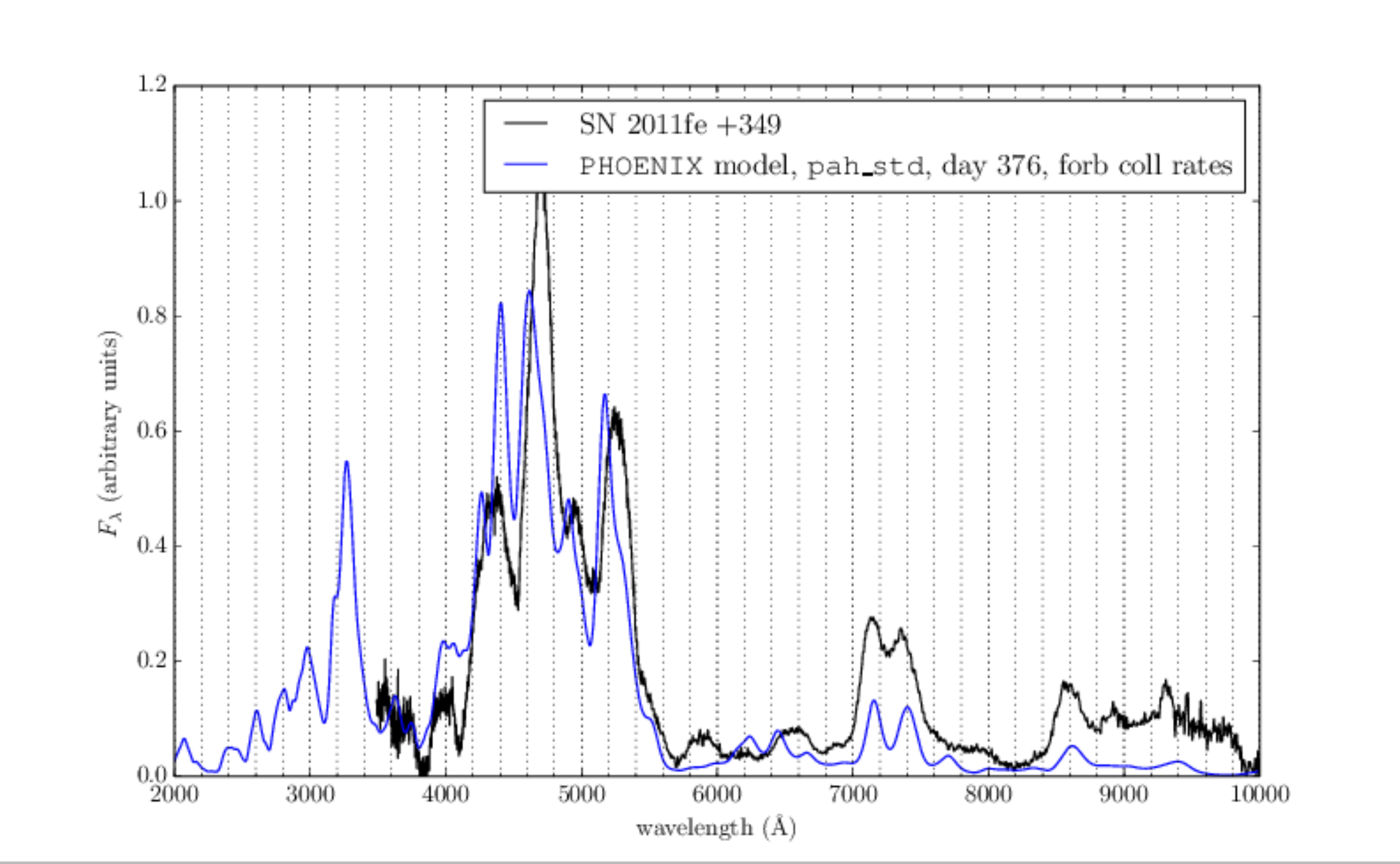}
    \caption{Synthetic spectrum of delayed-detonation model of \citet{dominguez01} at day 376 vs. SN~2011fe at day +349.}
    \label{fig:pah_std_d376_delta_t_vs_11fe_p349}
\end{figure*}

\begin{figure*}
\centering
    \includegraphics[scale=0.7]{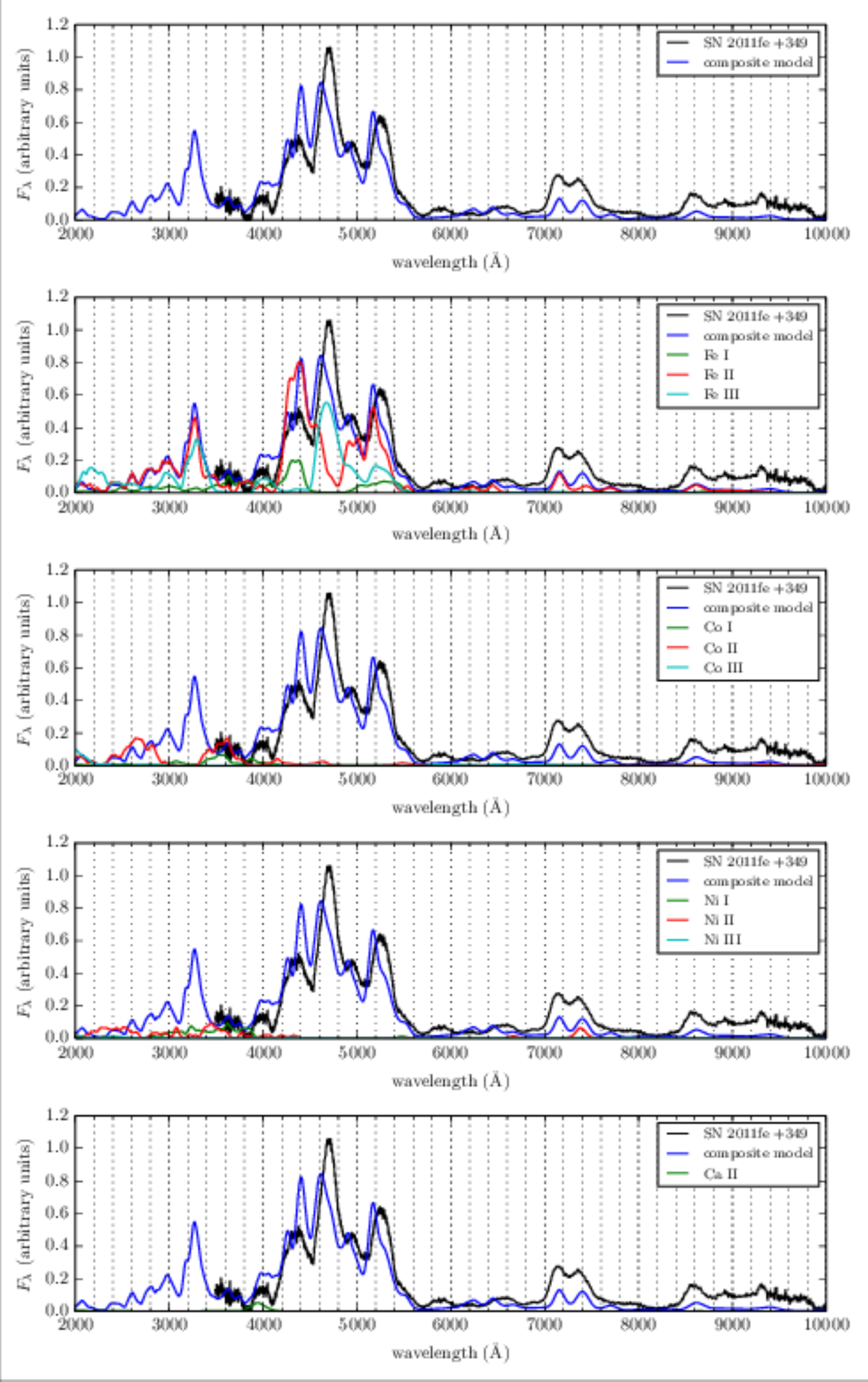}
    \caption{Single-ion spectra corresponding to the composite spectrum of the delayed-detonation model of \citet{dominguez01} at day 376, compared to SN~2011fe at day +349.}
    \label{fig:pah_std_d376_delta_t_forb_coll_vs_11fe_p349_single_ion_spectra}
\end{figure*}

The model and observations at day +349 are displayed in Figure~\ref{fig:pah_std_d376_delta_t_vs_11fe_p349}.
The corresponding single-ion spectra are shown in Figure~\ref{fig:pah_std_d376_delta_t_forb_coll_vs_11fe_p349_single_ion_spectra}.
The \ion{Ca}{2} H \& K doublet emission at 4000~\AA\ is of similar strength as at day +205, and it may still contribute to that feature in SN~2011fe, although the [\ion{Fe}{3}] appears to be stronger at that wavelength.

At this and later epochs the model begins to exhibit some problems.
A sharp emission feature forms in the near-UV in the synthetic spectrum which is not observed in SN~2011fe.
In addition, in the synthetic spectrum the emission at 4400~\AA\ (due mostly to \ion{Fe}{2}) is nearly as strong as the 4700~\AA\ feature (mostly \ion{Fe}{3}), while in the observed spectrum the latter remains considerably stronger.
This problem is likely not one of radiative transfer effects in the model, but rather one of atomic physics.
The recombination rate for ions scales with the free electron density $n_e$, which dilutes geometrically roughly as $t^{-3}$ \citep{de10a}.
Thus at these very late times the recombination time scale for, e.g., \ion{Fe}{2}, can be of the same order as the dynamical time scale, i.e., the age of the SN.
In this case, time-dependent effects of ion recombination can become influential on spectrum formation \citep{sollerman04,taubenberger15}.
In the calculations used to generate the above figures, we neglected time-dependence in both the radiation field and the ion populations: both are assumed to be in steady-state.
Assuming a steady-state radiation field is a valid approximation at late times --- since at most wavelengths the optical depths in the ejecta are low, the radiative transfer time scale is effectively the light-crossing time, which is many orders of magnitude shorter than the dynamical time scale.
Thus the radiation field equilibrates with the ejecta almost instantaneously at any given time $t$.
However, by assuming steady-state in the ion populations, we overestimate the rate of recombination from, e.g., \ion{Fe}{3} to \ion{Fe}{2}.
This manifests in synthetic spectra as \ion{Fe}{2} features which are too strong, as in Figure~\ref{fig:pah_std_d376_delta_t_vs_11fe_p349}.
Presumably the same pathology affects the day +578 synthetic spectrum more severely (see Figure~\ref{subsec:11fe_p578}).

Time-dependent effects in the ion populations may not be the only source of the discrepant features in the synthetic spectra at very late times.
The density profile of the ejecta in the explosion model also strongly affect $n_e$.
Thus the \ion{Fe}{2}-to-\ion{Fe}{3} population ratio may provide a constraint on the initial conditions of the explosion model.
For example, Figure~\ref{fig:pah_std_d376_delta_t_vs_11fe_p349} may indicate that the density of the iron-rich core of the model is too high, leading to an $n_e$ which is too high, inducing recombination from \ion{Fe}{3} to \ion{Fe}{2} too soon.

\subsection{The UV spectrum at day +360}
\label{subsec:11fe_p360}

\begin{figure*}
\centering
	\includegraphics[scale=0.5]{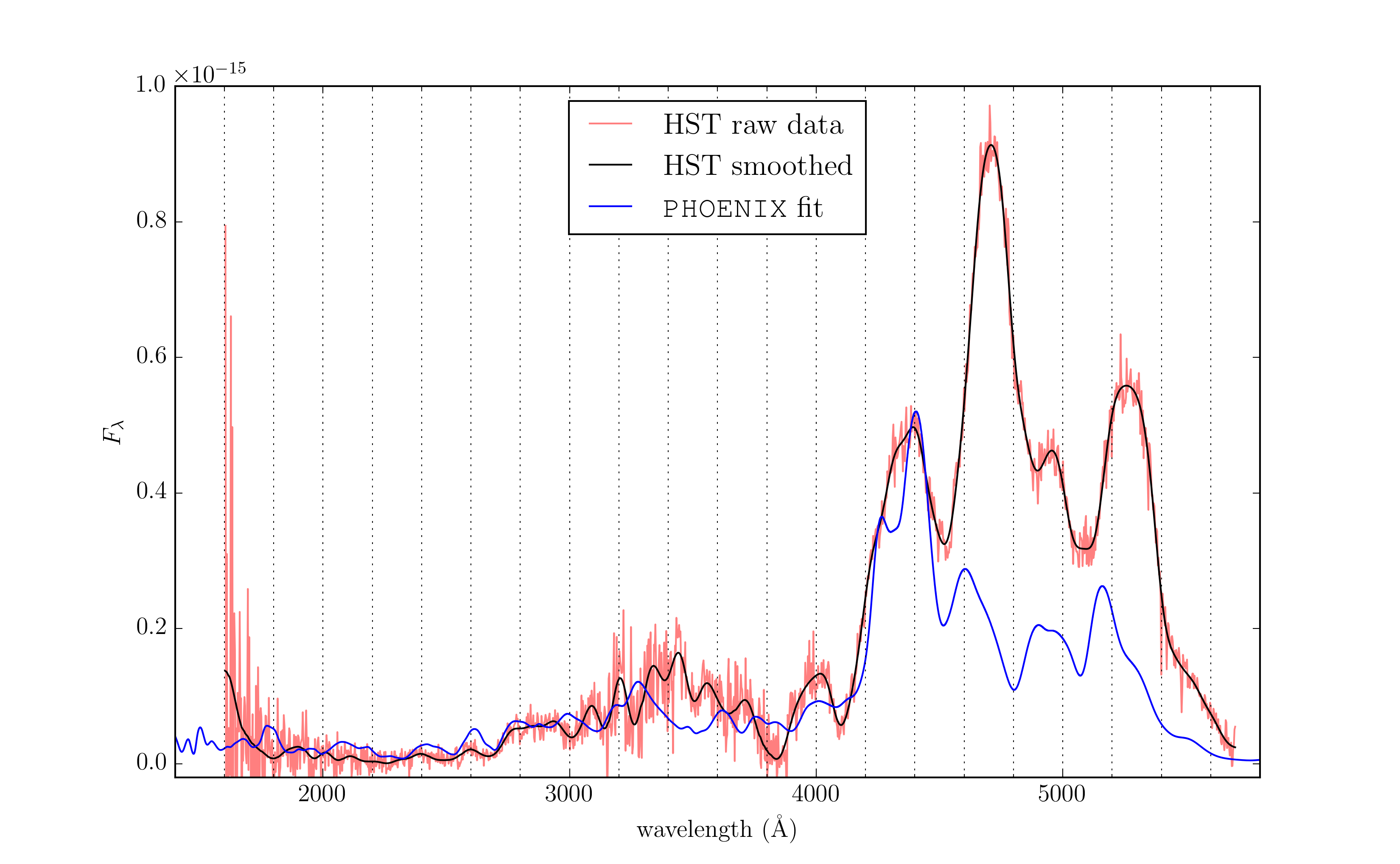}
    \caption{Synthetic spectrum of delayed-detonation model of
      \citet{dominguez01} at day 376, compared to SN~2011fe obtained
      with HST at day +360.}
    \label{fig:sn11fe_combined_hst_spectra_ft_smoothed_vs_pah_std_d376_fixed_tcor}
\end{figure*}

\begin{figure*}
\centering
	\includegraphics[scale=0.7]{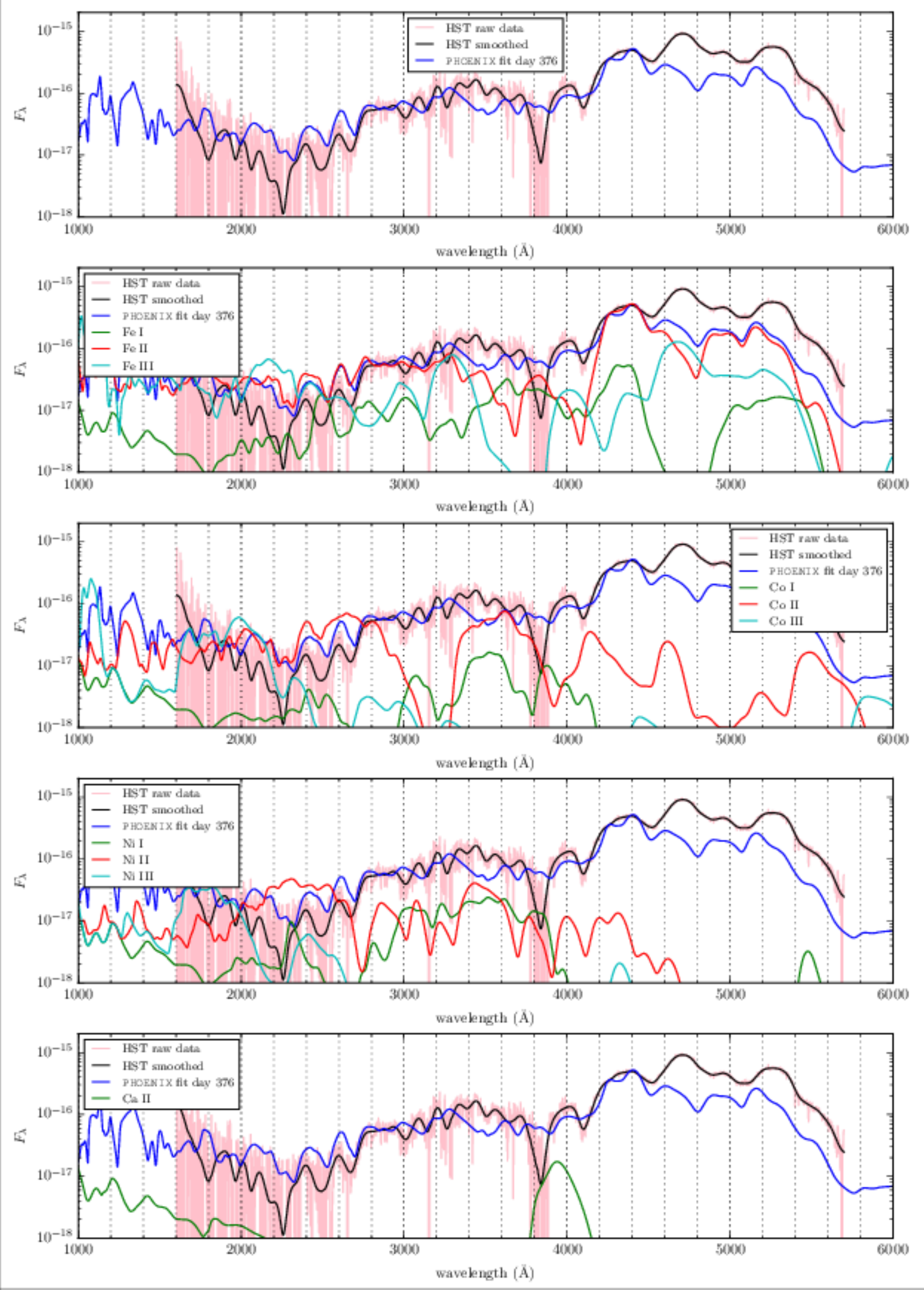}
    \caption{Single-ion spectra corresponding to the composite spectrum of the delayed-detonation model of \citet{dominguez01} at day 376, compared to SN~2011fe at day +360.}
    \label{fig:sn11fe_combined_hst_spectra_ft_smoothed_vs_pah_std_d376_fixed_tcor_single_ion}
\end{figure*}

The UV spectrum from \textit{HST} at day +360, as well as the best fitting spectrum from \texttt{PHOENIX}, are shown in Figure~\ref{fig:sn11fe_combined_hst_spectra_ft_smoothed_vs_pah_std_d376_fixed_tcor}.
The single-ion spectra are shown in Figure~\ref{fig:sn11fe_combined_hst_spectra_ft_smoothed_vs_pah_std_d376_fixed_tcor_single_ion}.
From these one finds that \ion{Fe}{2} is responsible for most of the spectral features in the UV at day +360, just as it was at day +100.
However, \ion{Fe}{3}, \ion{Co}{3}, and \ion{Ni}{3} all contribute significantly to the bluest portion of the spectrum as well.
The most interesting result, however, is that the \ion{Ca}{2} H \& K doublet continues to contribute significantly to the emission around 4000~\AA, despite being over 1~yr since explosion.
It seems, then, that the extreme strength of this line overcomes both the small total abundance of \ion{Ca}{2} in the ejecta, as well as the large amount of geometric dilution which accompanies a year of expansion.

A second notable feature of this result is that nearly all of the \ion{Fe}{2} features are permitted lines, not forbidden; the bluest forbidden lines for any ion in this version of the \texttt{PHOENIX} atomic database is about 3200~\AA.
Identifying these features is no simple feat, however, because \ion{Fe}{2} has thousands of lines between $\sim 1600 - 4000$~\AA.
Furthermore, the contributions from several other ions in the UV at late times, each with several thousands of lines themselves, are at some wavelengths of similar strength as \ion{Fe}{2}.
The convolution of all of these lines from different species renders the identification of individual features in the UV a difficult task.
However, even without identifying particular lines, we can nevertheless learn a great deal about the UV line forming region using other methods (see Figure~\ref{sec:opacity_late_times}).

\subsection{Day +578}
\label{subsec:11fe_p578}

\begin{figure*}
\centering
    \includegraphics[scale=0.7,trim=2cm .5cm .5cm .5cm,clip=true]{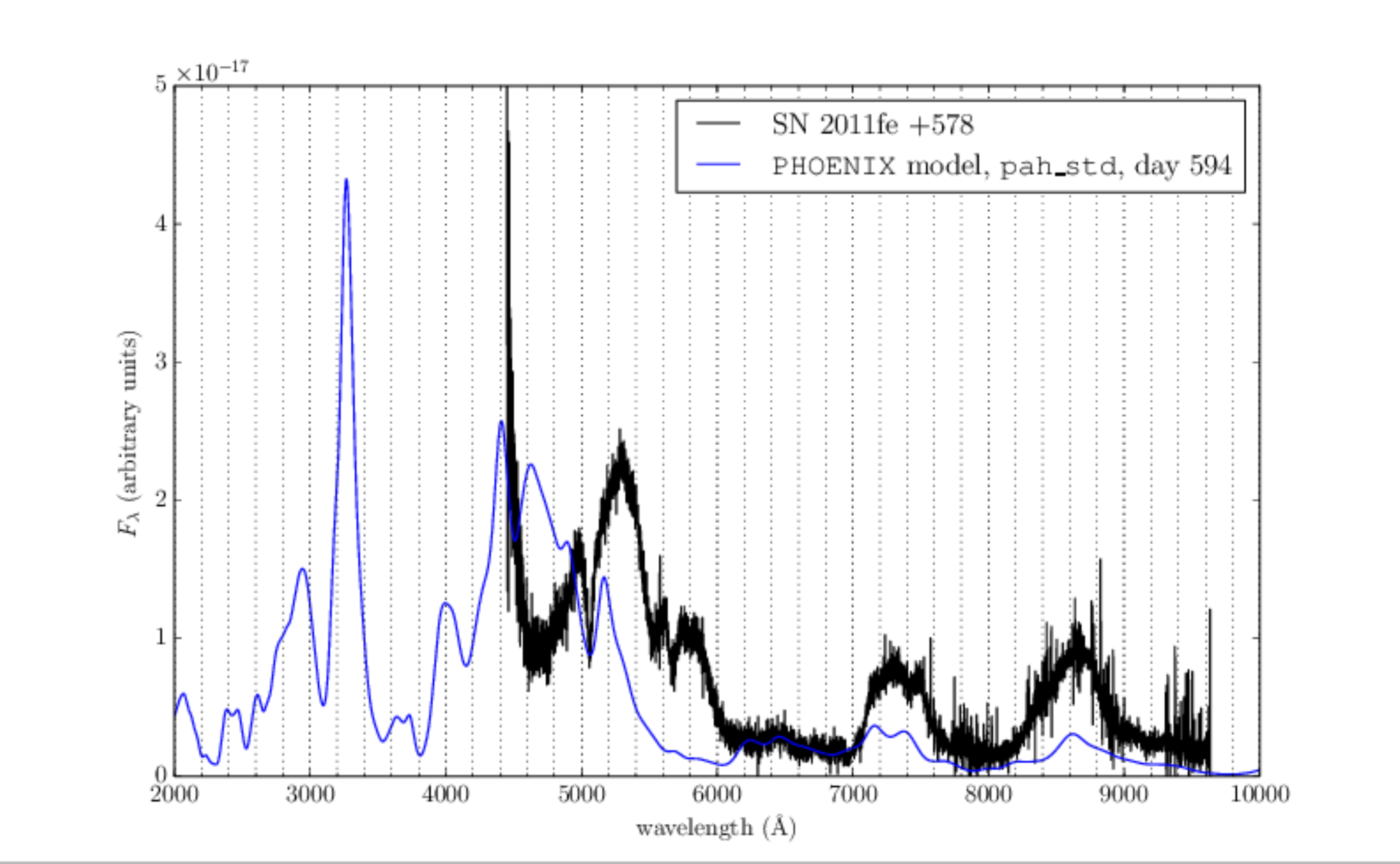}
    \caption{Synthetic spectrum of delayed-detonation model of \citet{dominguez01} at day 594 vs. SN~2011fe at day +578.}
    \label{fig:pah_std_d594_vs_11fe_p578}
\end{figure*}

\begin{figure*}
\centering
    \includegraphics[scale=0.7]{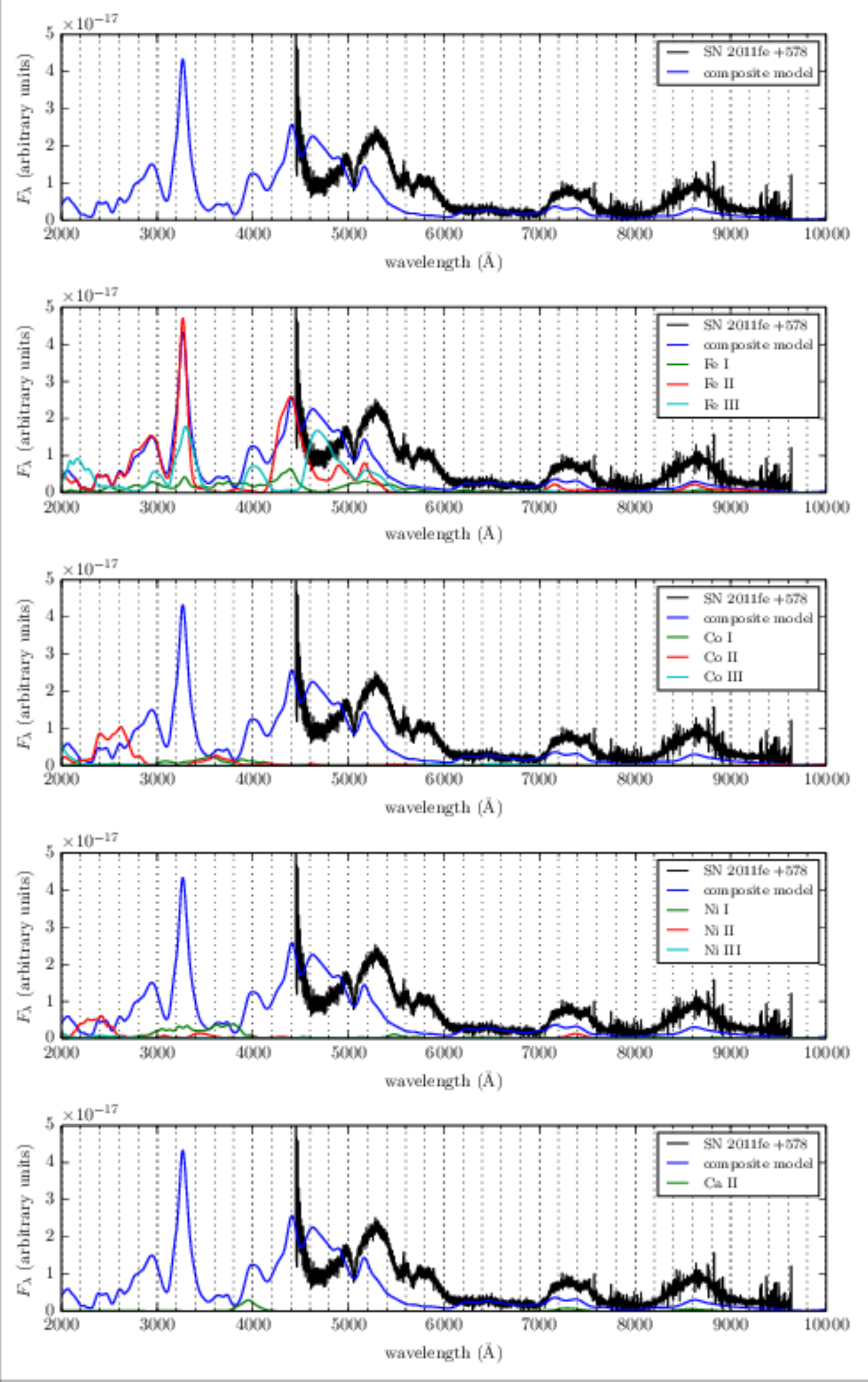}
    \caption{Single-ion spectra corresponding to the composite spectrum of the delayed-detonation model of \citet{dominguez01} at day 594, compared to SN~2011fe at day +578.}
    \label{fig:pah_std_d594_delta_t_forb_coll_vs_11fe_p578_single_ion_spectra}
\end{figure*}

Figure~\ref{fig:pah_std_d594_vs_11fe_p578} shows the observed and synthetic spectra of SN~2011fe at day +578.
The single-ion spectra are shown in Figure~\ref{fig:pah_std_d594_delta_t_forb_coll_vs_11fe_p578_single_ion_spectra}.
The fit of the \texttt{PHOENIX} synthetic spectrum to the observation is poor, and has resisted improvement even experimenting with a variety of different temperature-correction algorithms.
Possible culprits for this include time-dependent effects as discussed in \S\ref{subsec:11fe_p349}, as well as other physical processes which \texttt{PHOENIX} currently does not treat, including dielectric recombination and charge-exchange reactions.
However, analysis of this result nevertheless reveals some useful information.
For example, the unphysical spike in flux around 3250~\AA\ is due entirely to \ion{Fe}{2}, although to which line in particular is not clear.
In addition, the emission at 8600~\AA\, formerly produced by 
\ion{Ca}{2} IR3, has been replaced by [\ion{Fe}{2}]~$\lambda 8617$~\AA.
This is yet another example of a truly remarkable degeneracy among permitted lines and forbidden lines at similar wavelength, but which become active at very different times.

Although our model spectra predict the recombination to \ion{Fe}{2} too early, the event eventually does happen in SN~2011fe.
In particular, in the day +594 spectrum, the strong \ion{Fe}{3} emission peak at 4700~\AA\ has disappeared entirely, with only a handful of \ion{Fe}{2} features remaining.
Indeed, \citet{taubenberger15} and \citet{graham15} have tentatively identified features of \ion{Fe}{1} in $\sim 1000$~d spectrum of SN~2011fe, perhaps heralding a concurrent recombination transition from \ion{Fe}{2} to \ion{Fe}{1}.

\section{Velocity Shifts}

\begin{figure*}
\centering
    \includegraphics[scale=0.7]{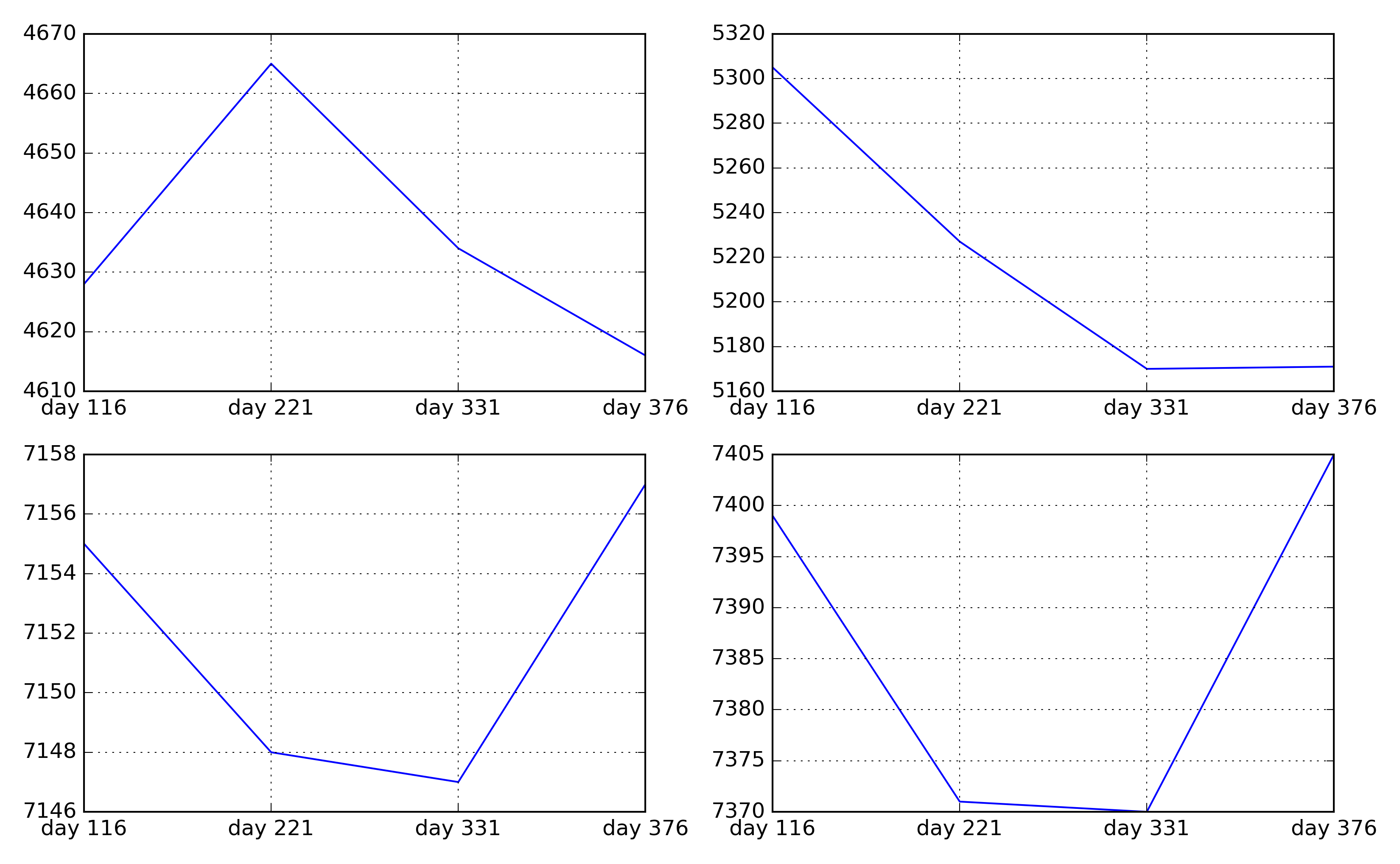}
    \caption{The position of the central wavelength peak of the
      4700~\AA, 5300~\AA, 7150~\AA, and 7400~\AA\ feature (clockwise
      from upper left) as a function of epoch.}
    \label{fig:feature_shifts}
\end{figure*}

In a recent paper, \citet{Black16} examined the wavelength shifts of
prominent features at late times in a series of SNe~Ia spectra,
including the spectra of SN~2011fe. They found, in particular, a
redward shift of of the prominent 4700~\AA feature with no signs of
the redward drift slowing down at epochs up to day
+400. Figure~\ref{fig:feature_shifts} shows that our models show no
such general trend. In fact, the 4700~\AA feature after a strong
redward shift, begins to move back to the blue. Since our models do
not show a strong fidelity with the observations it is difficult to
draw firm conclusions. \citet{Black16} suggest that the redward shift
is primarily caused by temporal variations due to lines of [Fe III]
and [Co II]. Since these lines are very temperature sensitive, this
could be indicative of our general model uncertainties.

\section{Opacity at late times}
\label{sec:opacity_late_times}

Our \texttt{PHOENIX} calculations have been fairly successful at reproducing the late-time optical and UV spectra of the normal SN~2011fe.
In addition, one of the great advantages of using first-principles codes such as \texttt{PHOENIX} is that one may glean a great deal of information about the underlying physics which drives the formation of the synthetic spectra.
For example, in Figure~\ref{fig:optical_depths_optical_p349} we show the optical depths along the $\mu = -1$ (radially inward) ray in the day +349 model whose spectrum was shown in Figure~\ref{fig:pah_std_d376_delta_t_vs_11fe_p349}.
For reference, the black dashed line shows $\tau = 1$, the division between optically thick and optically thin.
The dashed red line shows the optical depth due only to Thomson scattering; at early times this is the dominant opacity source and gives rise to the photosphere in SNe.
At late times, however, the geometric dilution of the free electron density $n_e$ leads to a very low Thomson scattering opacity, falling well below $\tau = 1$ even all the way to the center of the ejecta.
From this alone we may infer that there is likely very little continuum radiation present this late in a SN~Ia's lifetime, as reflected in the spectra.
Furthermore, at only a select few wavelengths -- mostly on the blue edge of the optical band -- does the optical depth reach $\tau = 1$ at all; at most wavelengths $\lambda \gtrsim 4500$~\AA, the ejecta are quite optically thin, in agreement with previous studies.
Our calculations therefore indicate that most of the optical spectrum at day +349 in SN~2011fe consists of blended emission features from collisionally excited, optically thin forbidden lines.

\begin{figure*}
	\centering
    \includegraphics[scale=0.6,trim=1.5cm .5cm .0cm .5cm,clip=true]{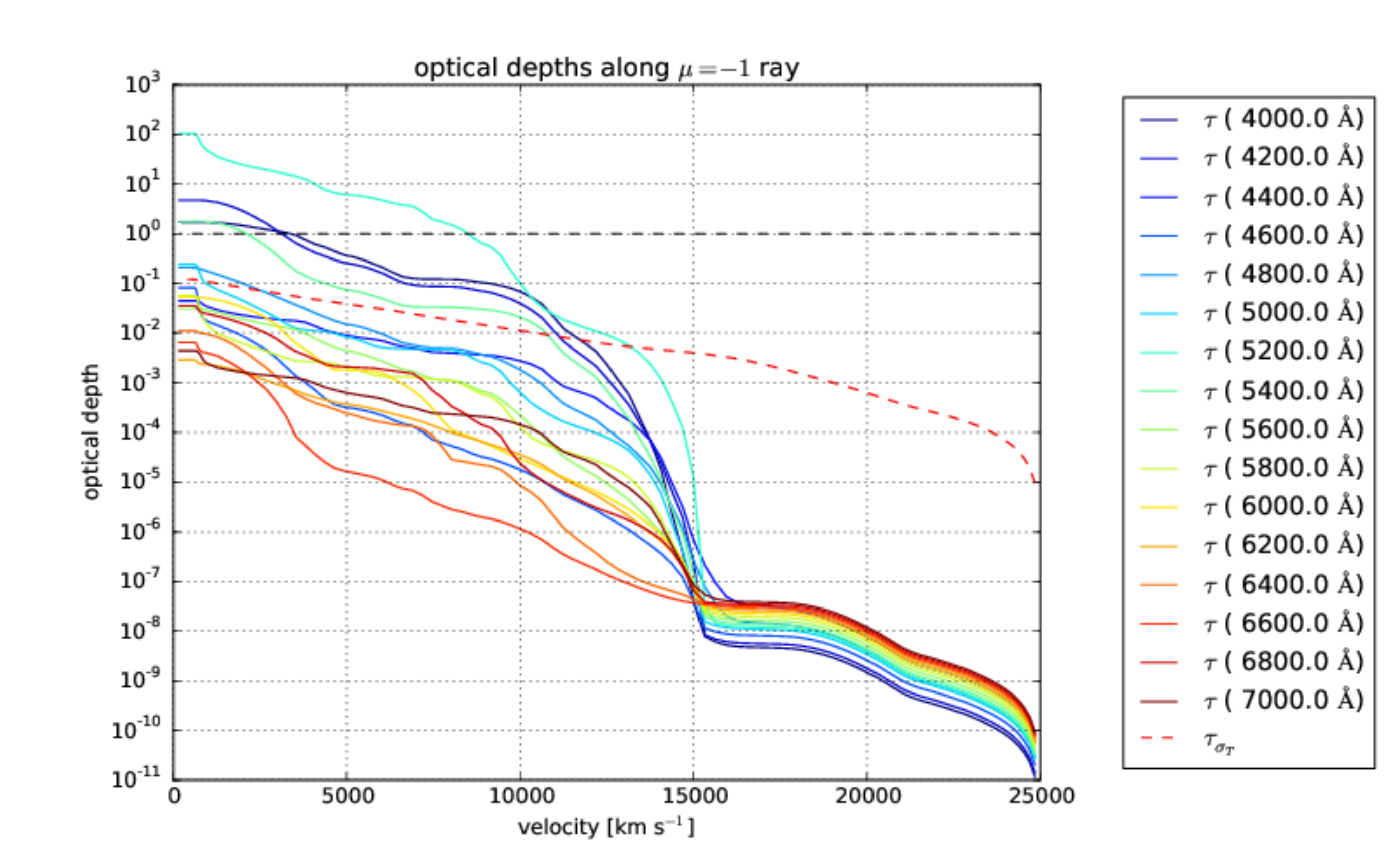}
    \caption{Optical depths at a collection of optical wavelengths in the day +100 model (cf. Figure~\ref{fig:pah_std_d116_delta_t_forb_lines_vs_no_forb_lines_vs_11fe_p100}).}
    \label{fig:optical_depths_optical_p100}
\end{figure*}

\begin{figure*}
\centering
    \includegraphics[scale=0.6,trim=1.5cm .5cm 0.0cm .5cm,clip=true]{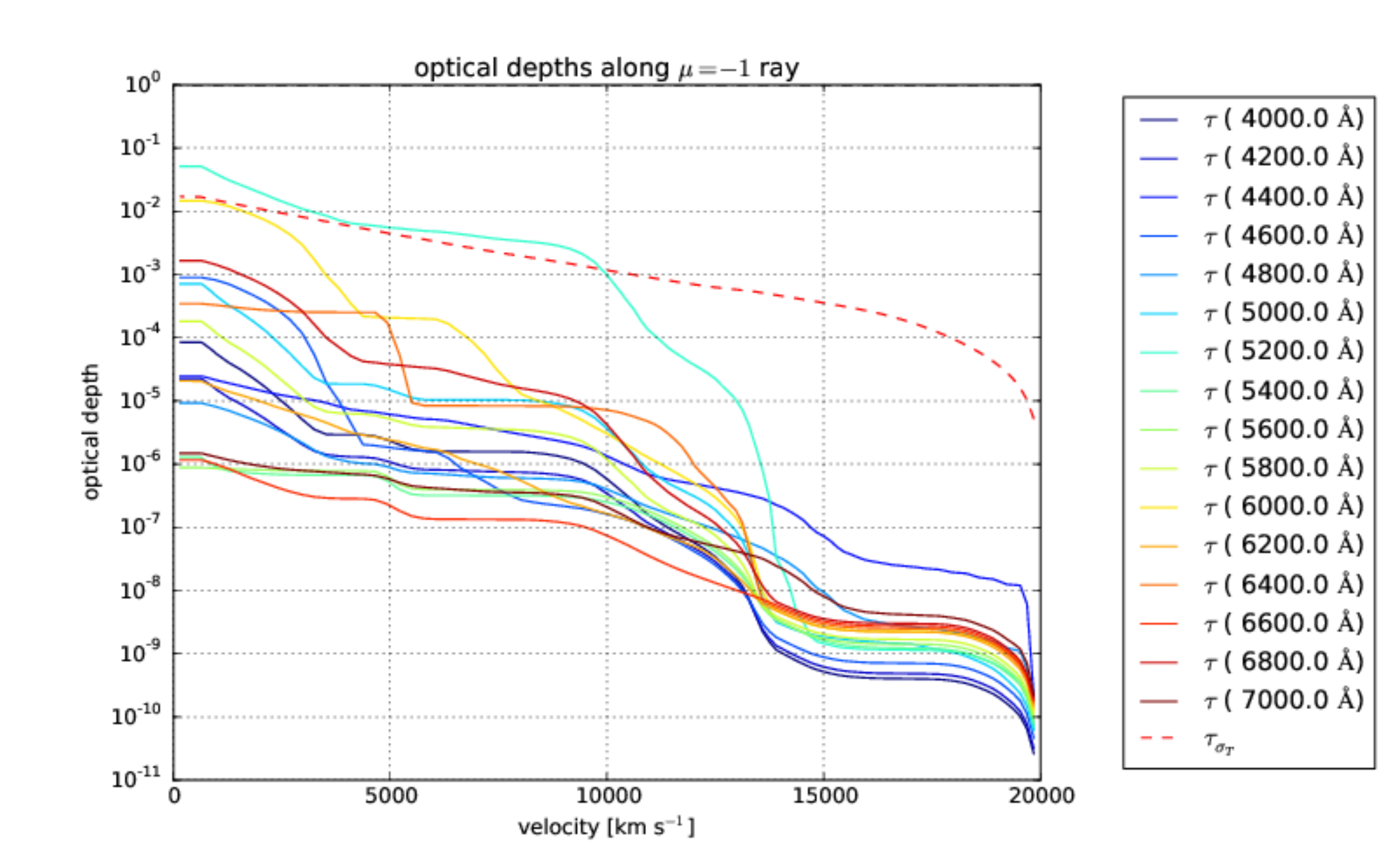}
    \caption{Optical depths at a collection of optical wavelengths in the day +349 model (cf. Figure~\ref{fig:pah_std_d376_delta_t_vs_11fe_p349}).}
    \label{fig:optical_depths_optical_p349}
\end{figure*}

The UV portion of the spectrum of SN~2011fe behaves entirely differently than the optical, however.
One may suspect as much simply by noticing the significant degree of structure and complexity in the observed UV spectrum (Figure~\ref{fig:sn11fe_combined_hst_spectra_ft_smoothed_no_model}).
These suspicions are confirmed by analogous calculations of optical depths at various UV wavelengths, shown in Figure~\ref{fig:optical_depths_UV_p349}.
The black and red dashed lines are the same as in Figure~\ref{fig:optical_depths_optical_p349}.
Unlike the optical band, however, most UV wavelengths are \emph{extremely} optically thick, with many reaching $\tau \sim 10^5$ at the center of the ejecta.
Another surprising result is that many UV wavelengths become optically thick at quite high velocity, crossing the $\tau = 1$ threshold at $v \sim 10 000 - 15000$~km~s$^{-1}$.
This result is corroborated by the presence of the emission component of the \ion{Ca}{2} H \& K doublet near 4000~\AA\ in Figure~\ref{fig:sn11fe_combined_hst_spectra_ft_smoothed_vs_pah_std_d376_fixed_tcor_single_ion}.

\begin{figure*}
\centering
    \includegraphics[scale=0.6,trim=1.5cm .5cm .0cm .5cm,clip=true]{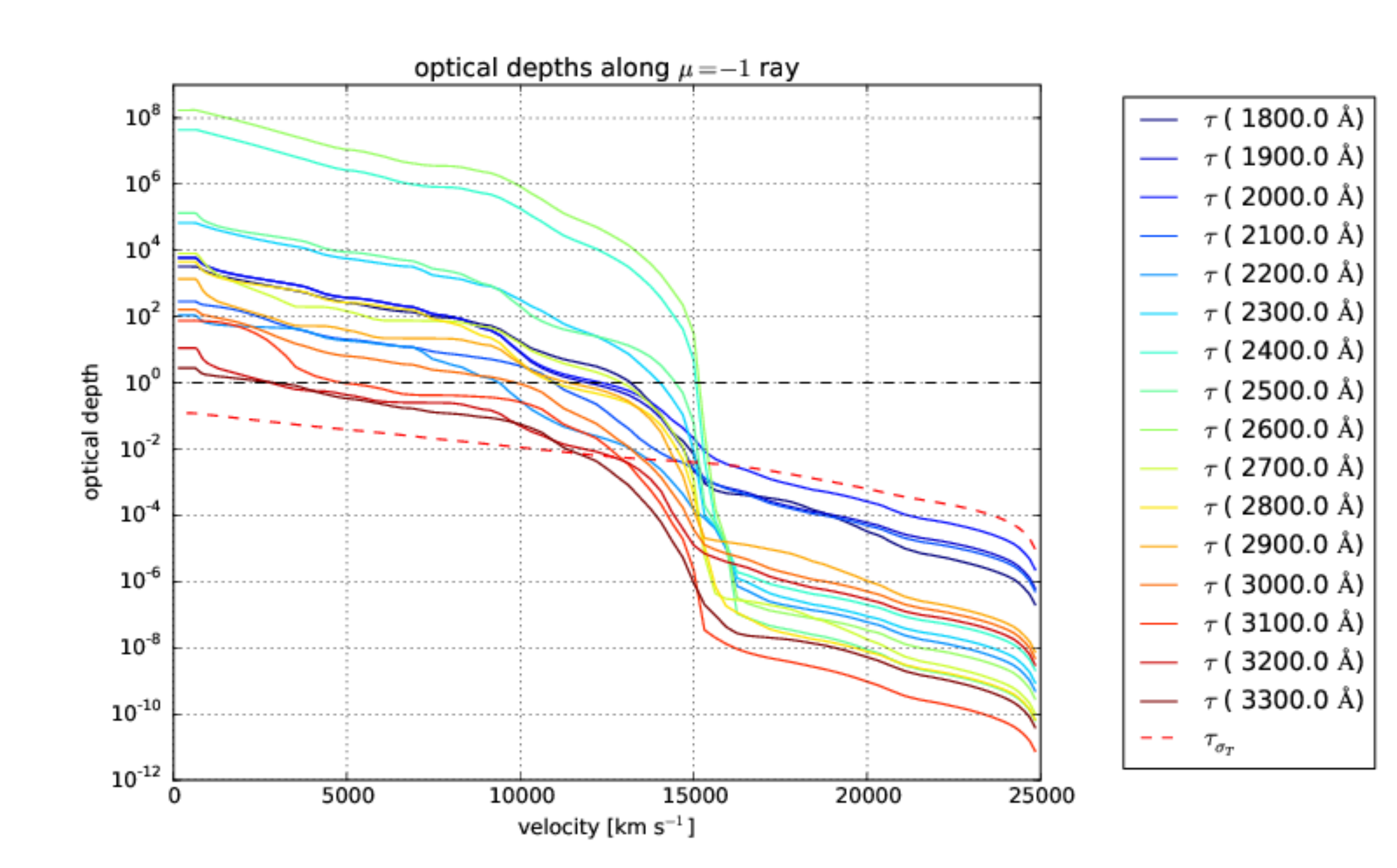}
    \caption{Optical depths at a collection of UV wavelengths in the day +100 model (cf. Figure~\ref{fig:pah_std_d116_delta_t_forb_lines_vs_no_forb_lines_vs_11fe_p100}).}
    \label{fig:optical_depths_UV_p100}
\end{figure*}

\begin{figure*}
\centering
    \includegraphics[scale=0.60,trim=1.5cm .5cm .0cm .5cm,clip=true]{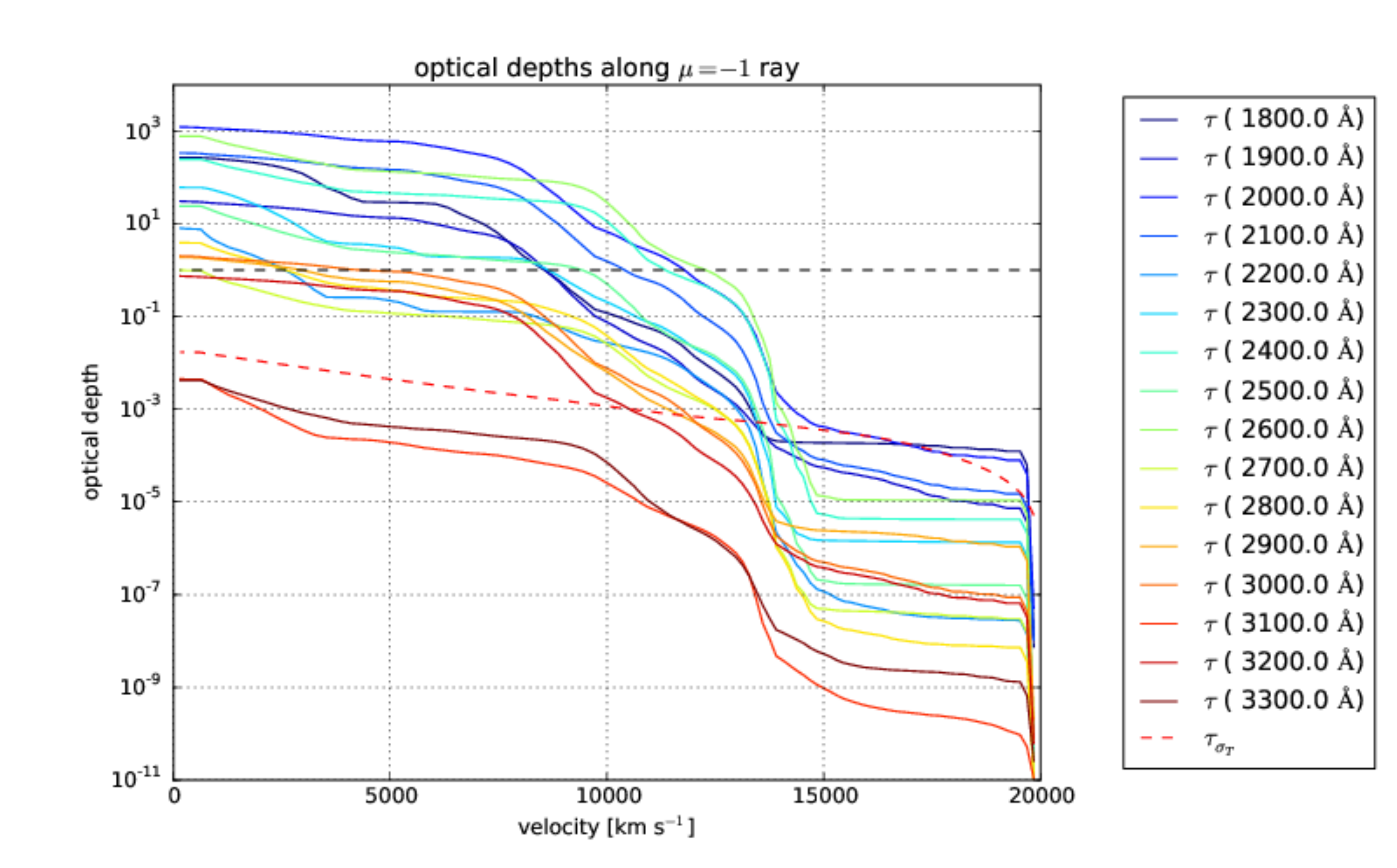}
    \caption{Optical depths at a collection of UV wavelengths in the day +349 model (cf. Figure~\ref{fig:pah_std_d376_delta_t_vs_11fe_p349}).}
    \label{fig:optical_depths_UV_p349}
\end{figure*}

If the UV remains as optically thick as Figure~\ref{fig:optical_depths_optical_p349} and Figure~\ref{fig:optical_depths_UV_p349} suggest, then it appears that the transition from the ``photospheric'' to the ``nebular'' phase in SNe~Ia is far more complex than expected.
Specifically, there are likely few or no forbidden emission lines which are active in the UV; this precludes the possibility of measuring asymmetric bulk motion of the inner regions of the SN ejecta, since the assumption behind such measurements is that the emission lines are optically thin and centered at the line rest wavelengths \citep{maeda10}.
On the other hand, since the UV is optically thick, the spectrum may consist of the same overlapping P~Cygni line scattering profiles which characterize maximum-light spectra of SNe; if this is the case we may be able to infer ejecta velocities of the iron-rich core of SN~2011fe by measuring the location of the absorption minima of the features in Figure~\ref{fig:sn11fe_combined_hst_spectra_ft_smoothed_no_model}.
However, as discussed in \S\ref{subsec:11fe_p360}, the multitude of UV lines, as well as likely blending among several atomic species, make this challenging.
If we entertain the possibility that some of the UV spectrum is forming at the UV photosphere at velocities of order $10 000$~km~s$^{-1}$, the rest wavelengths of such lines would still lie within a crowded space of UV and optical transitions of iron-peak elements.

The problem with the later fits may be related to time-dependent effects in the ionization of the gas, as the recombination time scale at late times is likely of order of the age of the SN.
By assuming steady-state in our calculations, we likely overestimated the recombination rate, leading to an overabundance of \ion{Fe}{2} with respect to \ion{Fe}{3}.
Re-computing these models with time-dependence in the NLTE rate equations is possible in principle, but the inherent ``noisiness'' of the root-finding algorithm we used to calculate the temperatures becomes amplified with each time step, resulting in a large amount of spurious temperature oscillations in the model at very late times, and a poorly fitting spectrum.

\section{Conclusions}

We extended \texttt{PHOENIX} to calculate radiative transfer models well into the late-time epochs of SNe~Ia, with an eye toward obtaining good fits to the high-quality optical and UV spectra of SN~2011fe.
Doing so required similar methods to those discussed in \citet{friesen14}, in particular using an alternative method to that of Uns\"old-Lucy for calculating the temperature structure of the gas, as well as accounting for the collisional and radiative rate data for forbidden lines, which behave quite differently than permitted lines.
The resulting synthetic spectra, ranging from +100 to +578 days post-maximum light, vary in degrees of fidelity to corresponding observed spectra of SN~2011fe, with the earlier epochs fitting quite well and the later epochs less so.
At day +100 we found that radiative transfer calculations which neglect forbidden lines and those which include them can produce remarkably similar optical spectra, but with quite different atomic species and combinations of lines forming the various features.
We found that, at least as late as day +360, permitted lines such as \ion{Ca}{2} H \& K and IR3 continue to influence spectrum formation in the optical, and permitted lines of \ion{Fe}{2} form much of the spectrum in the UV.
In addition, these models indicate that some emission features from permitted lines are replaced by other emission features of forbidden lines at nearly the same wavelength as the SN evolves.
For example, the emission from \ion{Ca}{2} H \& K at 4000~\AA\ is replaced around day +205 by [\ion{Fe}{3}]~$\lambda 4008$~\AA, and the emission from \ion{Ca}{2} IR3 at around 8600~\AA\ is replaced by [\ion{Fe}{2}]~$\lambda 8617$~\AA.

\section*{Acknowledgments}
The work has been supported in part by
support for programs
HST-GO-12948.004-A was provided by NASA through a grant from the
Space Telescope Science Institute, which is operated by the
Association of Universities for Research in Astronomy, Incorporated,
under NASA contract NAS5-26555.
This work was also supported in part by 
NSF grant AST-0707704, by NASA Grant NNX16AB25G and DOE Grant DE-SC0009956.
The work of EB was also supported in part by SFB 676, GRK 1354
from the DFG. R.J.F.\ gratefully acknowledges support from NASA grant
14-WPS14-0048, 
NSF grant AST-1518052, and the Alfred P.\ Sloan Foundation.
 
This research used resources of the
National Energy Research Scientific Computing Center (NERSC), which is
supported by the Office of Science of the U.S.  Department of Energy
under Contract No.  DE-AC02-05CH11231; and the H\"ochstleistungs
Rechenzentrum Nord (HLRN).  We thank both these institutions for a
generous allocation of computer time. 

\clearpage

\bibliography{bibliography}

\label{lastpage}

\end{document}